\documentclass[reprint,amsmath,amssymb,aps,prab]{revtex4-1}

\usepackage{graphicx}
\usepackage{dcolumn}
\usepackage{bm}

\begin{document}

\title{Gigahertz repetition rate thermionic electron gun concept}

\author{W.F. Toonen}
\author{X.F.D. Stragier}
\author{P.H.A. Mutsaers}
\author{O.J. Luiten}
\email[Electronic mail: ]{O.J.Luiten@tue.nl}
\affiliation{Department of Applied Physics, Coherence and Quantum Technology Group, Eindhoven University of Technology, P.O. Box 513, 5600 MB Eindhoven, the Netherlands}

\date{\today}

\begin{abstract}
We present a novel concept for the generation of gigahertz repetition rate high brightness electron bunches. A custom design $100$~kV thermionic gun provides a continuous electron beam, with the current determined by the filament size and temperature. A $1$~GHz rectangular RF cavity deflects the beam across a knife-edge, creating a pulsed beam. Adding a higher harmonic mode to this cavity results in a flattened magnetic field profile which increases the duty cycle to 30\%. Finally, a compression cavity induces a negative longitudinal velocity-time chirp in a bunch, initiating ballistic compression. Adding a higher harmonic mode to this cavity increases the linearity of this chirp and thus decreases the final bunch length. Charged particle simulations show that with a $0.15$~mm radius LaB\textsubscript{6} filament held at $1760$~K, this method can create $279$~fs, $3.0$~pC electron bunches with a radial rms core emittance of $0.089$~mm~mrad at a repetition rate of $1$~GHz.
\end{abstract}

\pacs{}

\maketitle

\section{\label{sec:intro}Introduction}
\begin{figure*}[t]
	\includegraphics[width=12.9cm]{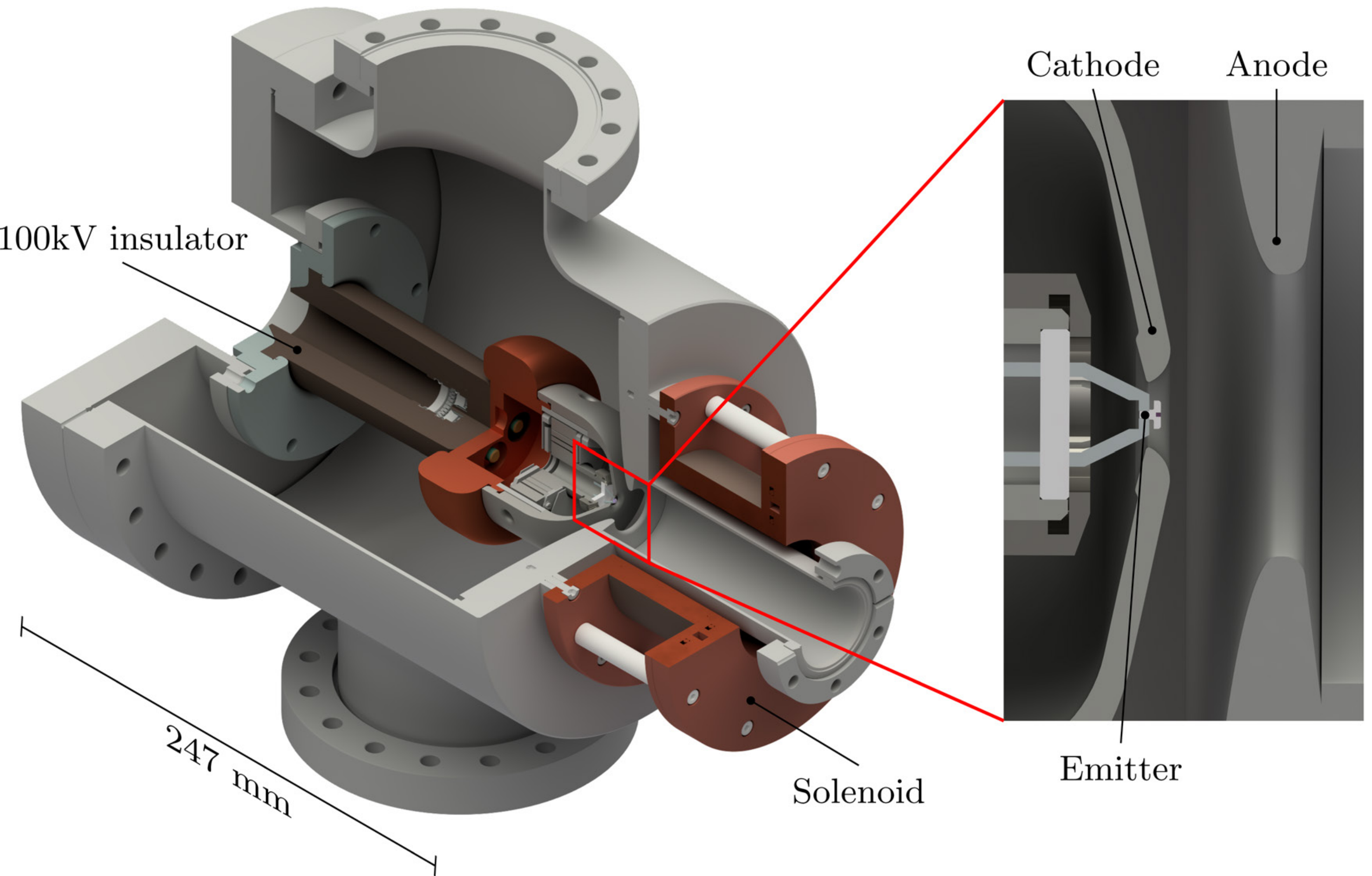}
	\caption{\label{fig:stp1}CAD model of the custom design thermionic gun DC accelerator. The 100 kV insulator is a Claymount R10 HV receptacle.}
\end{figure*}

High-brightness x-ray sources are used in a wide range of fields from chemistry, biology and medicine to material sciences, both in science and industry. Since the construction of the first synchrotron light source in 1986, they have provided researchers with an indispensable tool for non-destructive and high spatial resolution inspection of a wide variety of samples. 
With the growing demand for these high-brightness sources as well as the large size of the facilities that house them, spanning hundreds to thousands of meters, there is a desire for robust, compact and affordable x-ray light sources. One technique proposed to achieve this is in an Inverse Compton Scattering (ICS) scheme \cite{Graves2014}. This method requires an electron injector providing high-brightness, high charge, pulsed electron bunches that are subsequently accelerated in a radio-frequency (RF) accelerator. The bunch properties required at the point of electron-light interaction can be traced back to the properties of the bunches prior to injection into the RF accelerator. While current injectors are capable of providing a high \emph{peak} brightness electron beam, they are mainly limited in their \emph{average} brightness.
\\
\\
\begin{figure*}[t]	
	\includegraphics[width=17.2cm]{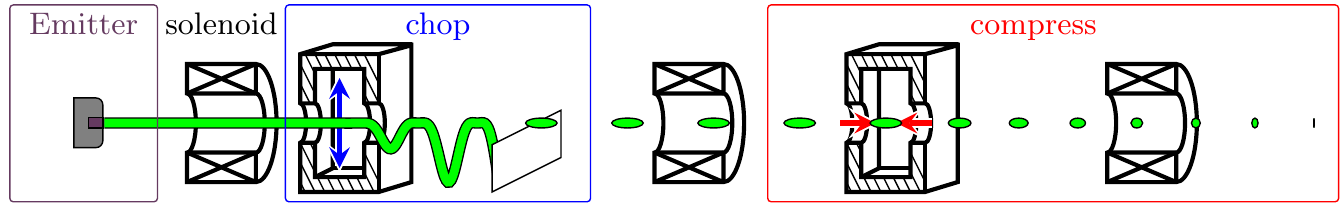}
	\caption{\label{fig:beamline} Schematic view of the complete thermionic electron gun. The blue and red arrows depict the electromagnetic force on the green electron beam due to the time-dependent magnetic and electric field respectively.}
\end{figure*}
Many avenues of research are being explored for the next-generation electron injector that will supply x-ray light sources with high charge, low emittance and high average current electron bunches. Since 1988, the common approach is to use an RF photoinjector, where a metal or semiconductor cathode is irradiated by a short, intense laser pulse \cite{Batchelor1988}. While these sources can achieve the highest brightness, metal cathodes have a relatively low quantum efficiency (QE), i.e.\ the number of electrons produced per incident photon. Semiconductor materials have a higher QE but in turn, require pressures below $10^{-9}$~mbar and have a limited lifetime of typically 1 to 100~hours, with performance often a trade-off between QE and durability. Advances in cathode development \cite{Filippetto2015,Dowel2010} and cavity design \cite{Musumeci2018} keep pushing these boundaries, with the APEX VHF-Gun a prime example \cite{Sannibale2012}. Future advances may include superconducting RF photoinjectors, capable of achieving even higher field gradients \cite{Vogel2018}. Alternatively, Cornell's DC photogun can deliver electron bunches with a bunch charge of $19$~pC and an emittance of $0.33$~mm~mrad at a $50$~MHz repetition rate or $77$~pC with an emittance of $0.72$~mm~mrad at $1.3$~GHz \cite{Bartnik2015,Gulliford2013}.
\\
Before photoinjectors became the standard, thermionic emission based guns were the default. Thermionic emission operates by heating the cathode so the thermal energy of the electrons can overcome the work function, essentially boiling off electrons from the cathode surface. The achievable beam current and emittance therefore heavily depend on the cathode material. While this method is less complex than a photoinjector, being able to operate in much relaxed vacuum conditions of $10^{-5}$ to $10^{-8}$~mbar and having thousands of hours of lifetime, common thermionic cathode materials such as tungsten or tantalum do not provide sufficient current density at a low enough emittance \cite{Jenkins1969}. However, with today's dispenser cathodes, as well as CeB\textsubscript{6} and LaB\textsubscript{6} crystals, ever lower work functions are reached. Researchers at SACLA have shown that a DC electron gun with a thermionic cathode can be successfully used as an injector for an x-ray free electron laser \cite{Togawa2007}. Thermionic cathodes in principle provide continuous electron beams, while acceleration to energies \textgreater$10$~MeV generally requires RF accelerators and thus a pulsed beam. At SACLA the method used is a beam chopper that can pick out a $2$~ns pulse at a $60$~Hz repetition rate. Another method often employed is through a gridded cathode, where a small grid in front of the cathode set at a potential at or higher than at the cathode. This way the amount of current reaching the anode can be regulated. The major downside of this method is the significant increase in beam emittance due to the presence of the grid.
Thermionic cathodes can also be used as the emitter in RF guns. Here, the gating is provided by the oscillating electric field, but this also induces increased energy spreads, as well as back-bombardment which results in cathode deterioration.\\
Finally, the injector at CEBAF uses a 100 keV beam chopper somewhat similar to the one proposed in this paper \cite{Abbott1994}. There, a square cavity with two orthogonal TM modes operating at the third subharmonic to their desired frequency sweeps an electron beam radially onto a circular aperture with three holes. This chops the beam into three beamlets, after which an identical second square cavity cancels the effects of the first one. Lens aberrations however lead to a minimum of $20$\% emittance growth.
\\
\\
In this paper, we present a novel approach to high-brightness, high repetition rate injectors based on thermionic emission and beam manipulation using RF cavities, as is schematically illustrated in Fig.~\ref{fig:beamline}. The thermionic electron gun, consisting of three separate stages, will first generate a continuous electron beam. Next, an RF deflection cavity containing a fundamental and higher harmonic mode will chop the beam into separate bunches. Finally, an RF bunching cavity, also running on two modes, will longitudinally compress the bunch to a sub-ps pulse length. The compressed bunch can then, for example, be injected into a booster linac capable of increasing the beam energy to \textgreater$10$~MeV. This injection occurs slightly before the point of maximum compression, so the beam will become relativistic at its minimal length. Hence, the point of maximum compression is the Point of Interest (PoI).\\
The first stage in the electron gun is a thermionic emitter, either a LaB\textsubscript{6} or CeB\textsubscript{6} crystal housed in a custom designed $100$~kV DC accelerator, generating a high average current continuous electron beam which is subsequently accelerated in the positive $z$-direction. With current technology these emitter crystals can be manufactured with an increasingly lower work function, resulting in a higher quality beam. Furthermore, these crystals can operate in a background pressure as high as $10^{-7}$~mbar while being capable of operating for thousands of hours, creating both a simple and robust system. Operating at $100$~kV instead of the more common $500$~kV reduces the requirement of high voltage insulation, leading to a smaller HV source and a smaller gun. The design of the gun is shown in Fig.~\ref{fig:stp1}. Construction of the DC accelerator is finished and initial testing has begun.
\\
In order to inject this beam into a booster linac, it has to satisfy certain entry conditions. First and foremost the beam should be pulsed with a maximum root-mean-square (rms) pulse duration of typically $0.1\pi$ to $0.2\pi$~radians of the booster linac RF phase \cite{Vretenar2013}. In the second part of the thermionic gun, the continuous beam is chopped into bunches by deflecting the beam onto a knife-edge using the on-axis transverse magnetic field of an RF deflection cavity. It has been shown that this chopping can be done without loss of beam quality \cite{Lassise2012}. However, the low duty cycle of this process defeats the purpose of using a continuous, high current density emitter. We will show that the addition of a second harmonic mode into the chopping cavity can increase the duty cycle to up to $0.6\pi$~radians phase angle of the fundamental mode, i.e.\ a 30\% duty cycle.\\ 
This increased duty cycle means the bunch is now too long to be injected into a booster linac. It is therefore required to compress the bunches by a factor of at least $6\frac{f_\textrm{boost}}{f_\textrm{chop}}$, with $f_\textrm{boost}$ the booster linac frequency and $f_\textrm{chop}$ the chopping cavity fundamental mode frequency. The third part of the thermionic gun will use velocity bunching to achieve this compression. By inducing a linear correlation between the longitudinal velocity and time, i.e.\ a negative $v_z$-$t$ chirp, the front of the bunch will overtake the rear, ballistically compressing the bunch over a drift space. In order to achieve this, we will use the on-axis longitudinal electric field of an RF bunching cavity. This method assumes that the change in electric field over time will be approximately linear. Near the zero-crossing of the field, this is certainly the case. Yet due to the increased duty cycle of the chopping process, the bunch will encounter distinctly non-linear parts of the field, resulting in insufficient acceleration and deceleration of respectively the front and rear end of the bunch as compared to the bunch center. This aberration decreases the compression ratio such that the bunches will be too long to enter the booster linac. The addition of a second harmonic mode into the cavity can account for these non-linearities, achieving sub-ps pulse lengths at the PoI. With this, the thermionic-based electron gun can deliver ultra-short, high quality, picocoulomb charge bunches at GHz repetition rates. An electrostatic deflector could then select a train of pulses for injection into the booster linac, allowing the electron gun to operate in either continuous wave or burst mode.\\
The second part of this paper (Sec.~\ref{sec:gun}) will describe the process of thermionic emission, leading to the design of the cathode and anode. In order to achieve tens of pC per bunch, the emission current is in the order of $100$~mA. At $100$~kV, this will require an HV generator capable of delivering a few tens of kW, combined with a beam block capable of dissipating it. In the first experiments, the gun will be operated at a power of $1$~kW where a $0.15$~mm radius emitter will deliver a $10$~mA average emission current to create a low emittance continuous beam. Appendix~\ref{sec:highcurrent} shows how higher average currents could be achieved.
\\
Sec.~\ref{sec:chopping} details the process of beam chopping, showing how a second harmonic theoretically improves the chopping efficiency. Charged particle tracking simulations are used for further optimization, demonstrating the increased effectiveness of the cavity operating at both the fundamental and second harmonic mode.\\
Sec.~\ref{sec:compress} describes the compression of the chopped beam, including a similar derivation for the higher harmonic mode. Simulation results will compare the bunch parameters with and without the addition of the second harmonic as well as the final parameters at the exit of the thermionic gun.
\section{\label{sec:gun}Thermionic Gun}
\subsection{Thermionic emission}
In thermionic emission the current density $J$ achieved by the emitter is described by the modified Richardson equation \cite{Dushman1923} 
\begin{equation}
J = A_g T^2 \exp(\frac{-(W-\Delta W)}{k_bT}),
\label{eq:RichEq}
\end{equation}
with $A_g$ a material constant called the Richardson constant, $T$ the crystal temperature, $k_b$ the Boltzmann constant, $W$ the work function of the material and $\Delta W$ the effective lowering of the work function due to the Schottky effect, given by \cite{Orloff2009}
\begin{equation}
\Delta W = \sqrt{\frac{e^3 E}{4\pi\varepsilon_0}}.
\label{eq:Schottky}
\end{equation}
Here, $e$ is the electron charge, $E$ the electric field at the cathode and $\varepsilon_0$ the vacuum permittivity. For LaB\textsubscript{6} the Richardson constant is $29$~A~cm$^{-2}$~K$^{-2}$, while work functions are reported between $2.3$ and $2.8 \:\text{eV}$ \cite{Lafferty1951,Pelletier1979}.
\\
If the electron bunches are to be injected into an accelerator-based light source, beam quality is a crucial aspect. In general, this quality is represented by the transverse normalized rms emittance \cite{Floettmann2003}
\begin{equation}
\varepsilon_{\textrm{n,rms}}=\gamma\beta\sqrt{\left\langle x^2\right\rangle\left\langle x'^2\right\rangle-\left\langle xx'\right\rangle^2},
\label{eq:rmsemi}
\end{equation}
with $\gamma$ the Lorentz factor, $\beta=v/c$ the velocity normalized to the speed of light $c$, the transverse position $x$ and the divergence in the paraxial approximation given by ${x'\approx v_x/v_z}$. As in x-ray generation schemes, the interaction happens mostly with the core of the phase space region of the beam \cite{Miltchev2005}. A better figure of merit therefore is the 90\% core emittance $\varepsilon^\textrm{core}$, which entails removing the outermost 10\% of the 4D transverse phase space prior to the calculation of the emittance.\\
At the cathode, the electrons are still being accelerated and this approximation is not yet valid. Instead, the initial emittance for a circular and uniform thermionic emitter can be calculated as the thermal emittance
\begin{equation}
\varepsilon_\textrm{n,rms}^\textrm{th} = \frac{r}{2}\sqrt{\frac{k_bT}{m_ec^2}},
\label{eq:thermalemittance}
\end{equation}
with $r$ the crystal radius and $m_e$ the electron rest mass. From Eqs.~(\ref{eq:RichEq}) and (\ref{eq:thermalemittance}) follows that, for a given current, a smaller radius with a higher temperature results in the highest beam quality. The lifetime of the emitter, however, is adversely affected by higher temperatures. For LaB\textsubscript{6}, limiting the temperature to below $1800\:\text{K}$ ensures a few thousands of operation hours; higher temperatures result in quickly decreasing lifetimes. Even if higher temperatures were possible, the current density cannot be arbitrarily increased. The limit is described by the Child-Langmuir law which states that as the current density increases, the electric field on the cathode caused by the space charge of the electrons will counteract the electric field of the DC accelerating structure. Current densities close to this limit are thus called space-charge limited. When operating in this regime, the electrons that escape the cathode are only accelerated slowly due to the smaller net electric field. This gives space-charge effects more time to influence the beam which may cause a significant growth in transverse emittance. This regime should therefore be avoided. For an infinite area cathode emitting electrons with zero velocity towards a parallel infinite area anode, the Child-Langmuir law is determined as \cite{Child1911}
\begin{equation}
J_\textrm{max}=\frac{4\varepsilon_0}{9}\sqrt{\frac{2e}{m_e}}\frac{V_0^{3/2}}{d^2},
\label{eq:Child}
\end{equation} 
with $V_0$ and $d$ the electric potential and the distance between the cathode and anode respectively.
In real electron sources, the cathode surface is small compared to the distance between cathode and anode, requiring the Child-Langmuir law to be multiplied by a factor $F$ depending on the ratio $r/d$ \cite{Togawa2007}. Limiting the cathode temperature to $1760$~K and taking a crystal radius of $0.15$~mm results in a thermal emittance of only  $0.04$~mm~mrad. With a current of $10$~mA and a chopping duty cycle of 30\%, a charge per bunch of $Q=3.0$~pC at a repetition rate of $1$~GHz is achievable. From Eqs.~(\ref{eq:RichEq}) and (\ref{eq:Schottky}) then follows that an electric field strength of approximately $10$~MV$/$m is required to reach this current. At an electric potential of $100$~kV, a gap of $10$~mm and the correction factor $F\approx 4.7$, the adjusted Child-Langmuir limit is approximately $245$~mA. This is well above the intended operating current of $10$~mA. Therefore, the electron source is not operating in the space-charge limited region.
\begin{figure}[b]
	\includegraphics[width=8.6cm]{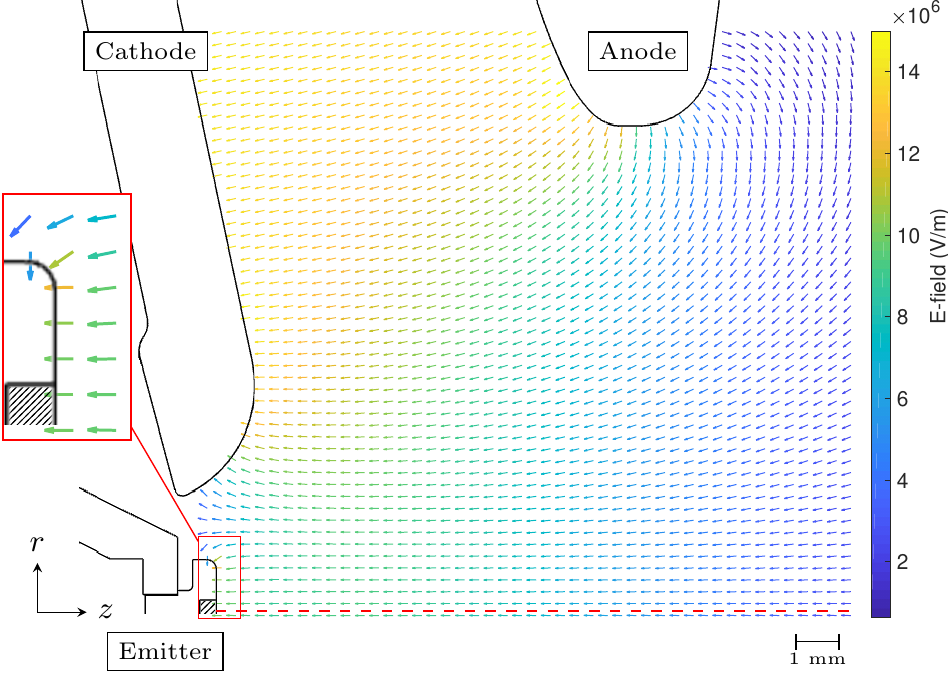}
	\caption{\label{fig:Efield} Electric field map between the anode and cathode. The field is rotationally symmetric with respect to the dashed red line, while the hatched area depicts the LaB\textsubscript{6} crystal.}
\end{figure}
\begin{figure*}[t]
	\includegraphics[width=8.6cm]{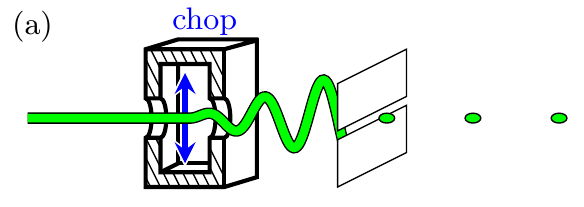}
	\includegraphics[width=8.6cm]{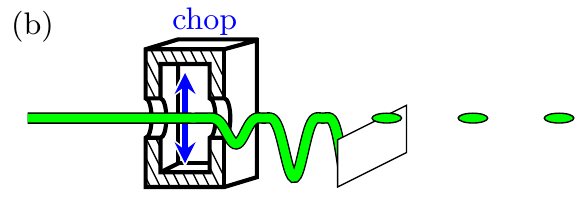}
	\caption{\label{fig:chop} (a) The principle of RF chopping. (b) The proposed method of chopping to increase the charge per pulse, using a fundamental mode, a higher harmonic mode and a constant magnetic field. The electron beam is shown in green and the force due to the oscillating magnetic field in blue.}
\end{figure*}
\subsection{Gun Design}
In order to achieve an electric field of $10$~MV$/$m at the emitter surface in a DC setup without breakdown, the custom cathode-anode assembly shown in Fig.~\ref{fig:stp1} has been designed. The geometry is optimized so that the electric field strength near the crystal is maximized, while simultaneously minimizing the increase everywhere else, as is shown in Fig.~\ref{fig:Efield}. The principal way this is done is by raising the cathode towards the anode hole while tapering the anode. This ensures minimal distance between the emitter and anode hole, while not decreasing the distance between cathode and anode, thus keeping the peak electric field strength low. Electrostatic simulations using \textsc{cst studio suite} \cite{CST} show that when $10$~MV$/$m at the emitter is reached, the peak field strength is approximately $15$~MV$/$m, near the edge of the emitter. This edge is a graphite ring with rounded edges that ensures no emission occurs on the side of the crystal. By making this graphite ring as large as $3$~mm across, the difference in the electric field strength between the crystal edge and center will be minimal. The size of the graphite ring, however, is currently limited by manufacturing restrictions and is set at $2$~mm across. With a field strength of $15$~MV$/$m, electron emission from the graphite will be over four orders of magnitude smaller than that of the crystal. Moreover, if an electrical breakdown should occur here, it shall be part of the thermionic current. This will only result in a temporary emitter current fluctuation and will not damage the accelerator structure.\\
After the electrons have been liberated from the crystal surface they first have to pick up speed and therefore spend a relatively large amount of time near the cathode. As the space charge forces rapidly expand the beam, the outer electrons will sample the off-axis non-linear fields, causing significant emittance growth. In order to counteract this, a magnetic solenoid lens is placed directly behind the anode to control the transverse beam size. However, the solenoid has a residual longitudinal magnetic field $B_{\textrm{res}}$ at the cathode, causing emittance growth in the beam due to an asymmetry between the azimuthal entrance and exit kick. This normalized rms magnetic emittance growth is given by \cite{Palmer1997}  
\begin{equation}
\varepsilon_\textrm{mag} = \frac{er^2\left|B_{\textrm{res}}\right|}{8m_ec}.
\label{eq:em_mag}
\end{equation} 
Since this emittance growth scales with $r^2B_{\textrm{res}}$, combined with the small size of the emitter as well as careful placement and tuning of the solenoid, the contribution of this effect should remain small. This will be investigated in Section~\ref{subsec:beamsim}.

\section{Beam chopping\label{sec:chopping}}
\subsection{RF cavity theory\label{subsec:cavtheory}}
The second part of the proposed setup is to chop the continuous beam of electrons into bunches with minimal loss of beam current, while maintaining beam quality.
\\
Chopping an electron beam can be done using RF cavities with an on-axis transverse magnetic field, as illustrated in Fig.~\ref{fig:chop}(a). The time-dependent transverse magnetic field deflects the beam periodically, after which an aperture blocks parts of the continuous beam, creating a train of ultra-short electron bunches \cite{Lassise2012thesis}. While this method does maintain beam quality, the downside is the limited duty cycle. Losing typically $\sim99$\% of the beam current on the aperture, a far greater initial current is required to achieve a decent charge per bunch.\\
In order to avoid this, the duty cycle can be increased by adding a higher harmonic mode to the cavity and choosing a different RF phase range in which electrons will pass the aperture. In the initial method that phase range is near the zero-crossing of the field. By adding a constant magnetic field, the electrons that pass the aperture are those that experienced the peak of the sinusoidal field. A higher harmonic mode can then be used to flatten that peak, as shown in Fig.~\ref{fig:chop}(b). Since chopping now only occurs on the top part of the beam, the aperture is replaced by a knife-edge.\\
Because the standing wave in an ideal cylindrical cavity is dependent on a Bessel function in the radial direction, a higher order mode is not an integer multiple of the fundamental frequency. This means a cylindrical cavity is inherently unfit for higher harmonic operation, increasing the difficulty of design.
However, the magnetic fields in a rectangular vacuum cavity operating in the TM\textsubscript{klm} mode are described by sinusoidal functions of position
\begin{align}
B_y&\left(x,y,z,t\right)=\nonumber\\&B_1\cos\left(\frac{k\pi}{a}x\right)\sin\left(\frac{l\pi}{b}y\right)\cos\left(\frac{m\pi}{d}z\right)\sin\left(\omega t\right),
\label{eq:recmodes}
\end{align}
with $k$, $l$ and $m$ the number of anti-nodes and $a$, $b$ and $d$ the cavity dimensions in respectively the $x$-, $y$-, and $z$-direction and $B_1$ the magnetic field amplitude. The propagation of the electrons is in the positive $z$-direction. This sinusoidal behavior means that with the correct cavity dimensions higher harmonics are possible. 
In order to determine which mode has to be added, we will look at the resulting time-dependent magnetic field along the $z$-axis. To achieve the greatest duty cycle without loss of beam quality, the magnetic field experienced by the electrons should remain constant for as long as possible. The ideal would be a rectangular wave. Approximating a rectangular wave with its Fourier series, however, requires many modes to reach suitable flatness. For experimental feasibility, we will only include two modes in the cavity.\\
With $k$ even, $l$ odd and $m=0$, the on-axis magnetic field of Eq.~(\ref{eq:recmodes}) for the first and the $\eta$-th order mode simplifies to
\begin{equation}
B_{y}^{1,\eta}\left(t\right) =B_1 \left(\sin\left(\omega_1 t+\phi_1\right)+\frac{1}{\zeta}\sin\left(\omega_{\eta} t+\phi_\eta\right)\right),
\end{equation} 
with $\omega_{\eta} = \eta\omega_1$, $\zeta = B_1 / B_{\eta}$ the amplitude scaling between the two modes and $\phi_{\eta}$ the phase of the $\eta$-th harmonic. To achieve a flat-top magnetic field profile around time $t=0$, all derivatives $d^nB_y/dt^n=0$ for $n\leq p$, with $p$ as large as possible, so that
\begin{equation}
B_{y}^{1,\eta}\left(t\right)\approx C+\mathcal{O}\left(t^{p+1}\right),
\label{eq:Taylor}
\end{equation}
with $C$ some constant. The higher $p$ is, the larger $t$ has to be in order for the field to deviate significantly from its value at $t=0$. Since $\phi_1$, $\phi_\eta$ and $\zeta$ can be controlled freely, the first three derivatives can be set to zero, resulting in
\begin{equation}
B_{y}^{1,\eta}\left(t\right) = B_1\left(\cos\left(\omega_1 t\right)-\frac{1}{\eta^2}\cos\left(\eta\omega_1 t\right)\right).
\label{eq:chopeq}
\end{equation}
In order to determine the optimal $\eta$, the fourth derivative at $t=0$ should be close to zero:
\begin{equation}
\frac{d^4B_{y}^{1,\eta}}{dt^4}\Bigr|_{t=0}=B_1\omega_1^4\left(1-\eta^2\right).
\label{eq:dd4}
\end{equation} 
Since integer $\eta>1$, the smallest absolute value is obtained for $\eta=2$, which means a second harmonic should be added to the cavity. Note that this is distinctly different from a square wave where the added term would be a third harmonic.
\begin{figure}[b]
	\includegraphics[width=8.6cm]{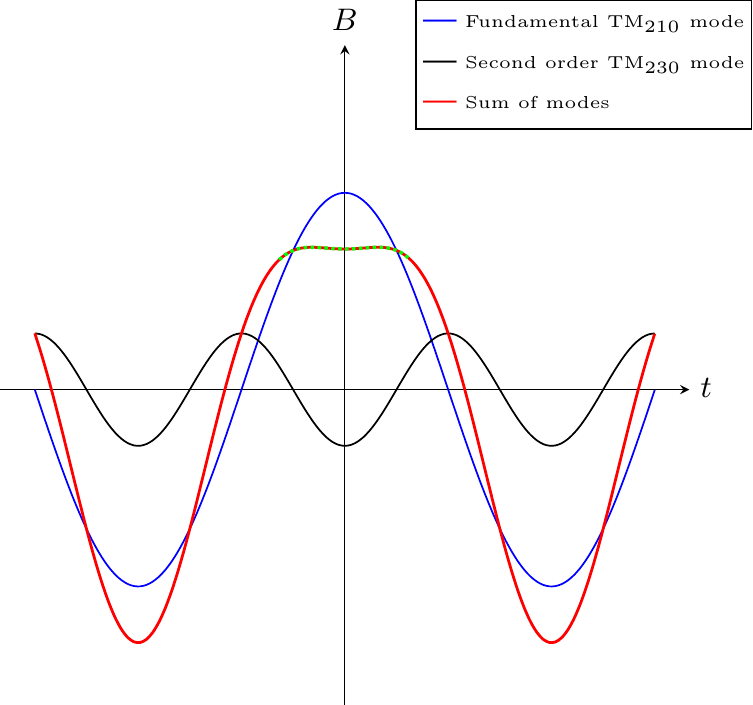}
	\caption{\label{fig:chopField} The time-dependent magnetic field in the center of the chopping cavity. The dashed green line represents the field probed by an electron bunch.}
\end{figure}
\begin{figure*}[t]
	\includegraphics[width=8.6cm]{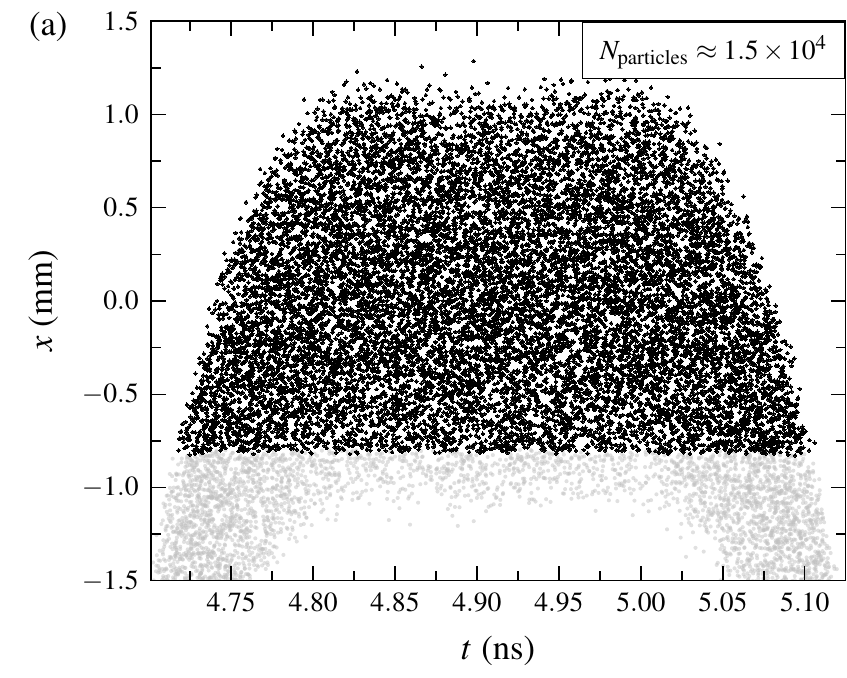}
	\hspace{\stretch{1}}
	\includegraphics[width=8.6cm]{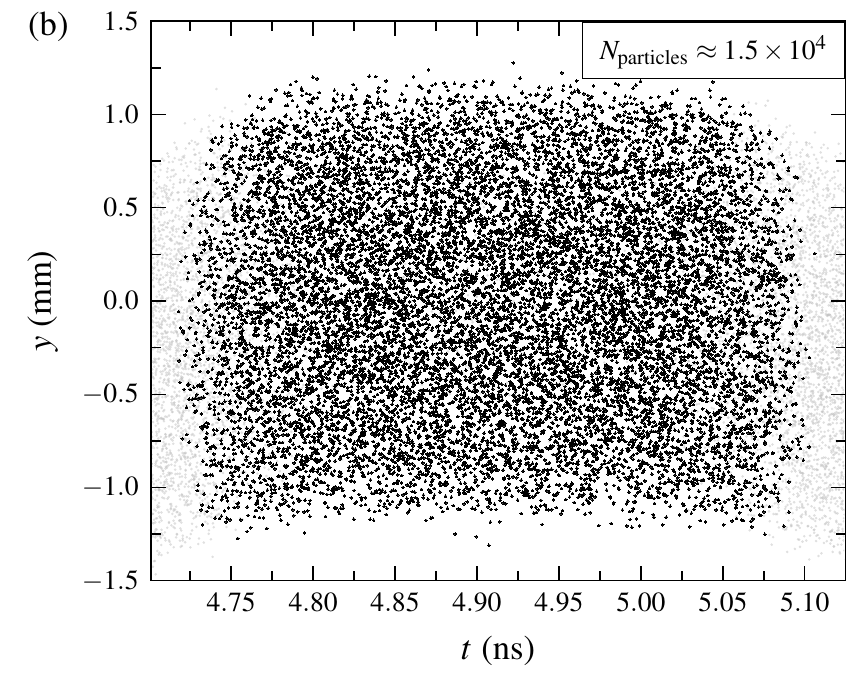}
	\caption{\label{fig:bunch} A single bunch after passing the knife-edge viewed (a) perpendicular and (b) parallel to the transverse chopping field. Total charge $Q=3.0$~pC. Each particle is a macro-particle representing approximately 1248 electrons. The grey particles were stopped by the knife-edge}
\end{figure*}
\\
A suitable combination of field modes for chopping is the TM\textsubscript{210} and TM\textsubscript{230} mode, of which the on-axis time-dependent magnetic field is shown in Fig.~\ref{fig:chopField}. In order for the latter mode to be the second harmonic of the first, the cavity dimensions must satisfy the relation
\begin{equation}
f_{230} = 2f_{210},
\label{eq:harmfreq}
\end{equation} 
with the resonance frequency given by
\begin{equation}
f_{klm}=\frac{c}{2}\sqrt{(\frac{k}{a})^2+(\frac{l}{b})^2+(\frac{m}{d})^2},
\label{eq:recfreq}
\end{equation} 
resulting in a ratio of 
\begin{equation}
a=b\sqrt{\frac{12}{5}}.
\label{eq:harmratio}
\end{equation} 
For a fundamental frequency of $1$~GHz the dimensions then are: $a=379$~mm, $b=245$~mm. The length of the cavity in the $z$-direction can still be chosen freely, as long as it does not cause other modes to be excited at or close to the fundamental or second harmonic frequency.
The fundamental frequency of $1$~GHz has been chosen for several reasons. With the higher order modes, the used frequencies are $f$ and $2f$. In the $1$ to $3$~GHz range the RF equipment is readily available. With a higher fundamental frequency, such as $3$~GHz, the required equipment is both more expensive and difficult to obtain. A lower frequency implies a lower repetition rate of the electron bunches, but also a proportional increase in charge per bunch. The downside of going to frequencies below $1$~GHz is the size of the cavities, which will exceed $0.5$~m at $425$~MHz, increasing both the size of the setup and the power required to drive the cavities.\\
By optimizing the power in the two modes it is possible to reach a duty cycle of approximately 30\%, leading to $3.0$~pC bunches at a bunch duration of $\sim96$~ps rms, which will be shown in the following section.
\subsection{\label{subsec:beamsim}Beam chopping simulations}
The entire beamline has been simulated using General Particle Tracer (\textsc{gpt}) \cite{GPT}. Since \textsc{gpt} can only simulate charged particle bunches, the continuous beam was simulated by using a $2$~ns long macro-bunch. This ensures that a large enough fraction in the center of this macro-bunch is not influenced by space charge effects due to the presence of a  front and back end. This fraction, effectively coming from a continuous beam, is the part that will eventually pass the chopping knife-edge.\\
Using the electric field distribution shown in Fig.~\ref{fig:Efield} as the accelerator field, the $100$~keV electron beam is generated at $t=0$ and $z=0$. As was demonstrated at SACLA, creating a thermionic source whose actual beam parameters agree with their theoretical value is experimentally achievable \cite{Togawa2007}. With a filament radius of $0.15$~mm, a temperature of $1760$~K and a field strength of $10$~MV/m, the $10$~mA beam is then generated with an emittance of $0.04$~mm mrad in both the $x$- and $y$-direction. The $100$~keV continuous electron beam passes through the first magnetic solenoid, which has a residual field on the cathode of approximately $8.3$~mT. Using Eqs.~(\ref{eq:thermalemittance}) and (\ref{eq:em_mag}), the total initial emittance is given by
\[
\varepsilon_\textrm{tot}=\sqrt{\varepsilon_\textrm{th}^2+\varepsilon_\textrm{mag}^2}\approx 0.0423\:\textrm{mm mrad}.
\label{eq:em_total}
\]
This result is similar to the $x$- and $y$-emittances of $0.0421$ and $0.0417$~mm~mrad respectively that follow from the charged particle simulation. While the simulation assumes proper alignment of the solenoid and the absence of any sources of jitter, this should not result in significant deviations.\\ The electron beam then enters the chopping cavity at $z=0.23$~m. 3D electromagnetic field maps of both the chopping and compression cavity were generated using \textsc{cst studio}, with the magnetic field of the chopping cavity oriented in the transverse $y$-direction. Details can be found in Appendix~\ref{sec:cstcavities}. The fundamental and second harmonic chopping modes operate at a $1$ and $2$~GHz frequency respectively, with separate control over the RF phases and amplitudes. Also added is a constant magnetic field in the $y$-direction. The strength of this field is set in such a way that the direction of propagation for the electrons of interest is not changed during the chopping of the beam, as is shown in Fig.~\ref{fig:chop}(b).\\
The amplitude of the fundamental mode is set at $B_1=1.0$~mT, with the second harmonic at $B_2=B_1/2.78$. Note that this relative amplitude of the second harmonic does not agree with the $1/\eta^2$ in Eq.~(\ref{eq:chopeq}). This is because the electrons are influenced by the integral of the field over their transit time through the cavity. This integration smooths out the field so that a higher relative strength can be used to increase the charge per pulse.
After passing the knife-edge at $z=0.57$~m, the electron beam now consists of $3.0$~pC bunches at a repetition rate of $1$~GHz. Since the chopping field is in the $y$-direction, the bunch oscillates and is chopped in the $x$-direction, as shown in Fig.~\ref{fig:bunch}(a). The grey particles indicate the electrons that hit the knife-edge, choosing the cut-off point so that there are no protrusions at the front and back of the bunch. The top of the bunch shows a small indent, which is caused by the aforementioned increased relative amplitude. Fig.~\ref{fig:bunch}(b) displays the same bunch, but since no chopping occurs in the $y$-direction the bunch shape will remain rectangular.
\section{\label{sec:compress}Pulsed beam compression}
\begin{figure}[t]
	\includegraphics[width=8.6cm]{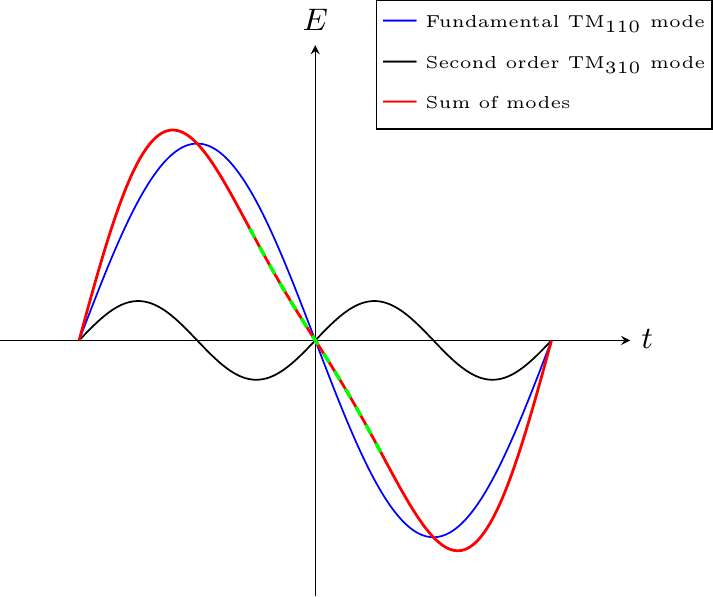}
	\caption{\label{fig:compress} The time-dependent electric field in the center of the compression cavity. The dashed green line represents the field probed by an electron bunch.}
\end{figure}

Like the cavity used for chopping, a rectangular RF cavity can be used for compression. With an on-axis longitudinal electric field, electrons passing through this cavity will experience a force solely in the propagation direction, causing them to be accelerated or decelerated. The amount of acceleration is determined by the integral of the electric field experienced and therefore depends upon the RF phase.
Due to the time-dependent field, different parts of an electron bunch will thus experience different amounts of acceleration. By setting the RF phase correctly, the center of the electron bunch will travel through the center of the cavity when the electric field goes through its zero-crossing. The electrons at the front of the bunch will then be decelerated while those at the rear will be accelerated, initiating a ballistic compression. 
This method of compression uses the fact that the electric field around the zero-crossing is approximately linear. For electron bunches at 30\% the length of one RF cycle this approximation is no longer valid. To allow longer bunch lengths to be compressed a higher harmonic mode can be used in order to shape the waveform.
\\
To achieve optimal compression, the resulting velocity distribution of the electrons should be linearly increasing from front to rear, i.e. a negative velocity-chirp. Assuming all electrons have the same longitudinal velocity, this means the change in momentum and thus the integrated electric field should be linear in time. This integrated field takes the shape of the moving average of the electric field, where the cavity length determines the interval of the integration. Therefore, if the electric field is sufficiently linear, the change in momentum and thus velocity will be as well.
Taking a similar approach as with the chopping cavity, the Taylor series of the electric field around $t=0$ should then be of the form
\begin{equation}
E_z^{1,\eta}\left(t\right)\approx Ct+\mathcal{O}\left(t^{p+1}\right),
\end{equation}
with $C<0$.
This means setting the zeroth, second and third derivative to zero. Doing so will also set the fourth derivative to zero, so $\eta$ is used to minimize the fifth derivative, leading to
\begin{equation}
E_{z}^{1,2}\left(t\right) =E_1 \left(-\sin\left(\omega_1 t\right)+\frac{1}{8}\sin\left(2\omega_{1} t\right)\right),
\end{equation} 
which is schematically illustrated in Fig.~\ref{fig:compress}. 
\\
In general, the modes suitable for compression are the TM\textsubscript{2$\kappa$+1,2$\lambda$+1,0} modes with $\kappa$ and $\lambda$ non-negative integers. The fundamental mode at frequency $f$ is TM\textsubscript{110}, where the ratio of dimensions for the double frequency $2f$ mode can be determined similar to the chopping cavity using Eqs.~(\ref{eq:harmfreq}) and (\ref{eq:recfreq}), resulting in
\begin{equation}
a=b\sqrt{\frac{5}{3}}
\label{eq:harmratio2}
\end{equation} 
for the TM\textsubscript{110} and TM\textsubscript{310} mode. For a $1$~GHz cavity, $a=245$~mm and $b=190$~mm.
\subsection{Bunch compression simulations}
\begin{figure}[t]
	\includegraphics[width=8.6cm]{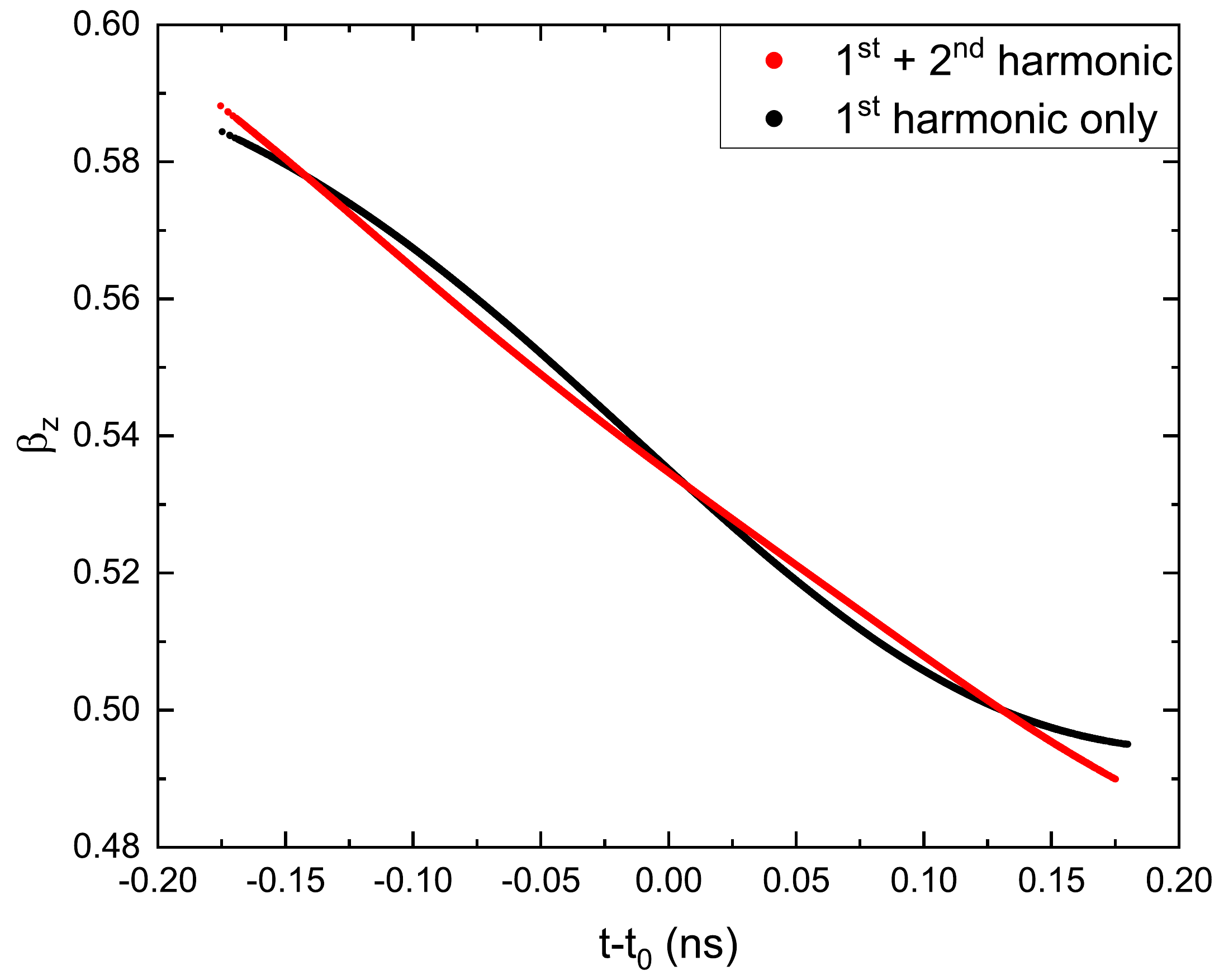}%
	\caption{\label{fig:t-zprime}\textsc{gpt} simulated longitudinal phase space $(t,\beta_z)$ of a bunch directly behind the compression cavity.}%
\end{figure}
\begin{figure}[t]
	\includegraphics[width=8.6cm]{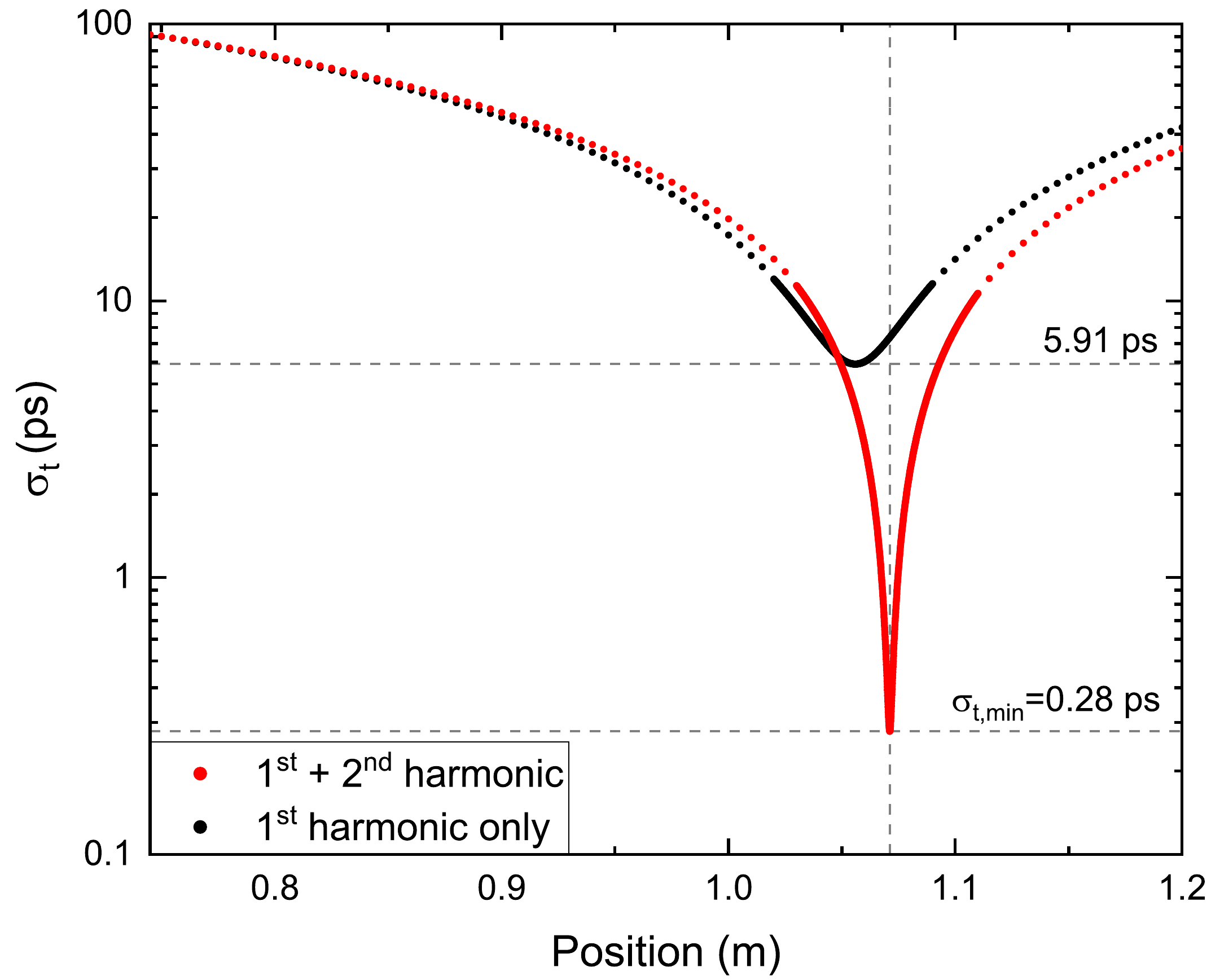}
		\caption{\label{fig:beamend_t}RMS temporal bunch length during compression. Initial pulse length is $95.57$~ps.}
	\includegraphics[width=8.6cm]{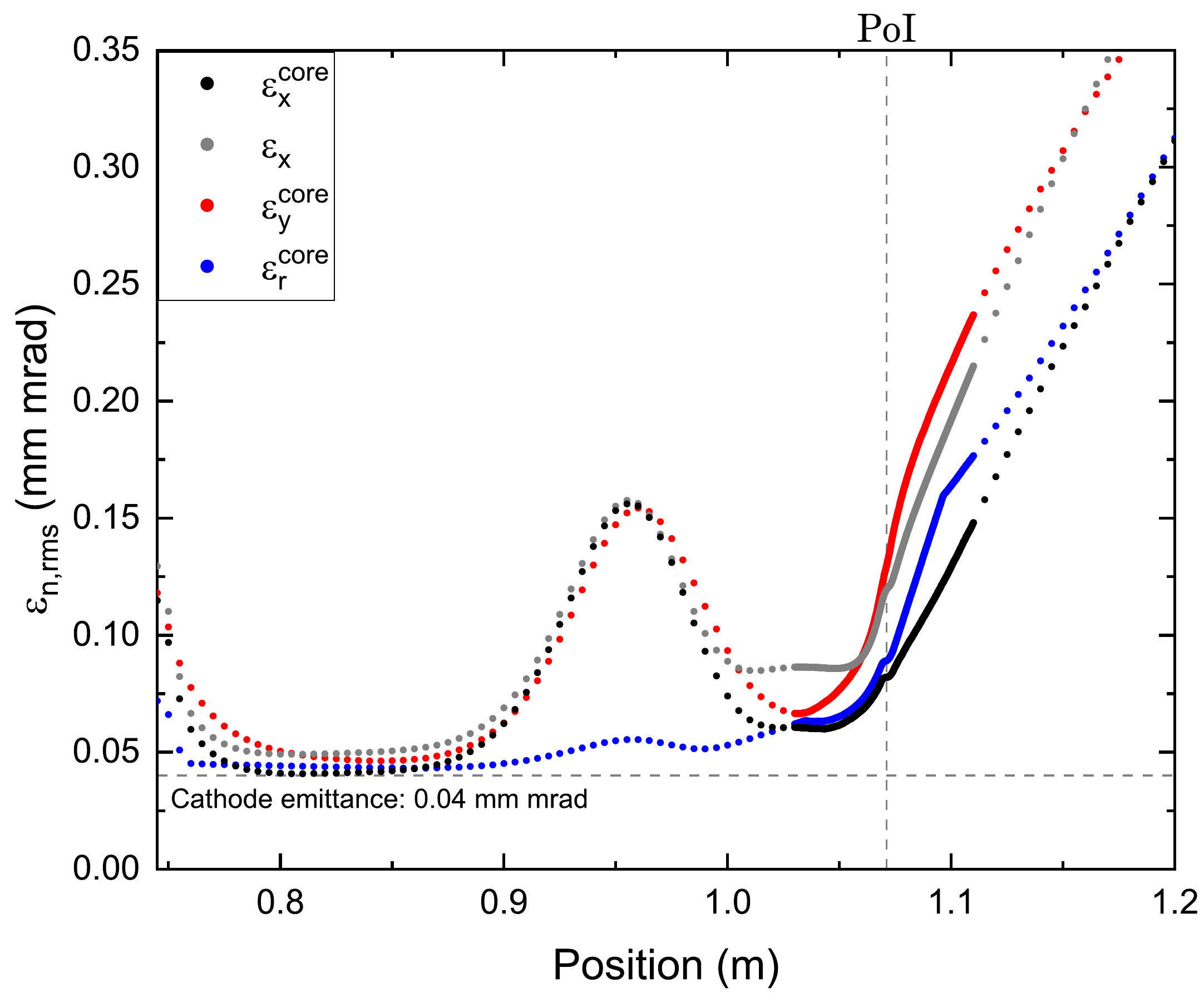}
		\caption{\label{fig:beamend_emi}Emittances of the bunch at the end of the beamline. The bunches pass through the PoI at $z=1.07$~m.}
\end{figure}

After the chopping cavity and knife-edge, the bunches drift towards a magnetic solenoid that controls the beam radius, after which it finally enters the RF compression cavity at $z=0.73$~m. Here, the $1$~GHz and $2$~GHz mode initiate the bunch compression. In the absence of space charge effects, a perfectly linear relationship between $t$ and $\beta_z$ would result in a minimal compressed pulse length, directly related to the energy spread of the bunch prior to compression. Fig.~\ref{fig:t-zprime} shows that adding a second harmonic to the cavity significantly increases the linearity of this relationship. As with the chopping cavity, the optimal mode parameters are different from their theoretical value, as they can somewhat account for non-linearities in the bunch, such as the initial energy spread and space-charge effects. Using \textsc{gpt} for optimization, the mode parameters are $\left\lbrace \phi_1,\phi_{\eta},\zeta\right\rbrace =\left\lbrace 0.91\pi,0.04\pi,3.41\right\rbrace$ instead of the theoretical $\left\lbrace \pi,0,8\right\rbrace$, with $E_1=7.06$~MV/m. The large difference for $\zeta$ is mainly because the cavity length is not taken into account in the theoretical value. As a longer cavity results in more averaging of the field, greater deviations from a linear electric field can still result in a linear integrated field, which is ultimately what leads to ballistic compression.
\\
As shown in Fig.~\ref{fig:beamend_t}, after drifting for about $0.34$~m and through another solenoid at $z=0.97$~m, the electron bunches will pass through the PoI. Comparing this result with that of the optimal compression using only the first harmonic shows a significant decrease in rms pulse length, creating $3.0$~pC bunches as short as $279$~fs at a $1$~GHz repetition rate. Fig.~\ref{fig:beamend_t} also shows a slightly longer focal length for the higher harmonic compression cavity. As the second harmonic is in anti-phase to the first harmonic, the effective electric field is weaker than without the second harmonic, causing the difference. This is also visible in Fig.~\ref{fig:t-zprime}, where the slope near $t=0$ in the longitudinal phase space is slightly lower for the higher harmonic compression.
\\
The emittance from Eq.~(\ref{eq:rmsemi}) can be used to compare the bunch quality during compression. However, since the chopped bunch was part of the top of a modified sine wave, it is asymmetrically shaped. Added to this is that the range of RF phases that will allow electrons to pass the knife-edge is changed, depending on how the knife-edge is positioned. If a greater range is allowed, the bunch in Fig.~\ref{fig:bunch}(a) will be more "banana-shaped", with the front and rear of the bunch extending further in the negative $x$-direction, as depicted by the grey particles. These parts have a significant effect on the emittance but a lesser effect on x-ray generation, which the 90\% core emittance takes into account. Furthermore, beamline elements such as a magnetic solenoid introduce large correlations between the transverse positions and velocities of the electrons, causing a large but temporary emittance growth. By looking at the radial emittance
\[
\varepsilon_{r,\textrm{n,rms}}=\gamma\beta\sqrt{\varepsilon_{x,\textrm{rms}} \varepsilon_{y,\textrm{rms}}-\left|\left\langle xy\right\rangle\left\langle x'y'\right\rangle-\left\langle xy'\right\rangle\left\langle x'y\right\rangle\right|} 
\]
the contributions to this growth can be reduced.   
Fig.~\ref{fig:beamend_emi} shows the core $x$-, $y$- and $r$-emittances after going through the compression cavity at $z=0.73$~m. At position $z=0.81$~m the core emittance has increased to $0.041$~mm~mrad. Comparing this to the emittance of $0.049$~mm~mrad shows that the majority of emittance growth is due to the outer 10\% of the phase space.\\
Comparing Figs.~\ref{fig:beamend_t} and \ref{fig:beamend_emi} shows a significant increase in transverse emittance around the PoI. This is due to space-charge effects that start to dominate the short electron bunch, but this increase can be reduced by decreasing the amount of compression or decreasing the charge per pulse. In short, at the PoI the thermionic electron gun provides $279$~fs electron bunches with a $0.089$~mm~mrad radial core emittance at a $1$~GHz repetition rate.

\subsection{\label{subsec:cavfeas}Cavity Feasibility}
As mentioned in Section~\ref{subsec:cavtheory}, since the chopping and the compression cavity both operate at the fundamental and the second harmonic mode, a rectangular cavity is preferred. Research into power efficient RF cavity design, however, has mainly been done for pillbox cavities \cite{Verhoeven2018}. This is accomplished either through modifying the geometry of the cavity or through the introduction of dielectric material into the cavity, where the cylindrical symmetry somewhat simplifies the design process. For a rectangular cavity, such options have not yet been explored, as far as we know.\\
\textsc{cst studio} simulations show that for a regular rectangular copper compression cavity, the total power required to drive both the fundamental and second harmonic mode to the field amplitudes used in this paper is approximately $539$~W. For the chopper cavity, $425$~W is required to reach a peak field of $B_1=1$~mT. No optimization has been performed on these geometries, leading to these moderately high powers. Still, CW solid-state amplifiers are commercially available in this power and frequency range and should pose no problem.
\section{Conclusions}
With the need for a next-generation high-repetition-rate, high-brightness electron injector, we have presented a design concept for a $100$~keV pulsed electron gun based on the chopping and compression of a continuous beam from a LaB\textsubscript{6} cathode.\\
The use of an RF cavity for beam chopping allows the electron gun to reach GHz repetition rates without severe degradation of beam quality. Adding a higher harmonic mode, the duty cycle of this process is greatly increased, reaching a high charge per bunch relative to the initial current, resulting in $279$~fs, $0.089$~mm~mrad, $3.0$~pC charge electron bunches at $1$~GHz. With both the high repetition rate and an increased bunch length, compression to sub-ps pulse lengths can be achieved with an RF compression cavity whose compression fields have been linearized by adding a higher harmonic mode.
Altogether, the thermionic electron gun is a compact setup capable of producing an electron beam at a wide range of currents, with a fast chopping process designed to retain as much charge as possible and a compression process designed to match the chopping cavity, all the while keeping beam emittance low. Once proof of principle is given, the setup will be upgraded to higher average currents in order to provide the charge, quality and repetition rate required for applications such as ICS operation.
\\
Currently, the construction of the first stage of the thermionic gun, the DC accelerator, has completed and initial experiments are being performed.
\begin{acknowledgments}
The authors would like to thank Eddy Rietman and Harry van Doorn for their invaluable technical support in the mechanical and electrical design of the thermionic gun. This research is part of the High Tech Systems and Materials programme of the Netherlands Organisation for Scientific Research (NWO-AES) and is supported by ASML.
\end{acknowledgments}

\appendix
\section{High current operation\label{sec:highcurrent}}
As Section~\ref{sec:chopping} has shown, a $10$~mA continuous electron beam sent through the $1$~GHz chopping cavity will result in a charge of $Q=3.0$~pC per bunch. Since the 30\% duty cycle of this process cannot be arbitrarily increased, higher bunch charges can only be reached by increasing the current. However, the current density from Eq.~(\ref{eq:RichEq}) has already been maximized: $A_g$ and $W$ are determined by the emitter material, $\Delta W$ is limited by the peak electric field in the gun and $T$ cannot be increased without severely affecting the lifetime of the filament. As such, the only method of increasing current is to increase the emitter radius. Several things have to be taken into account for this greater radius.\\
First, the charge per bunch will increase quadratically with the emitter size, while the thermal emittance will increase linearly. For example, going from $r=0.15$~mm to $r=0.75$~mm will increase $Q$ to $75$~pC and the thermal emittance to $0.20$~mm~mrad. The new radius also changes the multiplication factor for Eq.~(\ref{eq:Child}) from $F\approx 4.7$ to $F\approx 2.8$. Recalculating the currents from Eqs.~(\ref{eq:RichEq}) and (\ref{eq:Child}) comes to an emitted current of $250$~mA. Meaning the emission current is still over ten times smaller than the space-charge limited current of $3.6$~A.\\
Second is the radial component of the DC accelerator field near the emitter. The graphite ring surrounding the crystal is not only used to prevent emission from the side of the crystal but also decreases the radial component of the electric field near the crystal. It is therefore important that the emitter crystal has sufficient distance to the emitter edge so that only the longitudinal fields are probed by the electrons.\\
Thirdly, the increased radius means the emittance contribution of the solenoid increases as well. Care has to be taken that this does not have too great an impact on the final bunch emittance. If necessary, a bucking coil could be introduced into the system.\\
Finally, the increased beam width means deviations in both beam chopping and compression will occur. While this might be beneficial for the charge per bunch, it comes at a cost of an increased emittance and a decreased compression ratio. Additional focusing or changing the chopping cavity parameters can provide some compensation for this. Some changes in the cavity design, such as the cavity hole size, might be necessary.\\
All things considered, scaling towards higher currents mainly involves changing peripheral devices, such as an HV power supply capable of delivering $25$~kW and a beam block capable of dissipating that power. Once high current operation is achieved, switching between high and low currents involves little more than switching to a different filament.
\section{\label{sec:cstcavities}RF cavity design}
\begin{figure*}[t]
	\includegraphics[width=17.2cm]{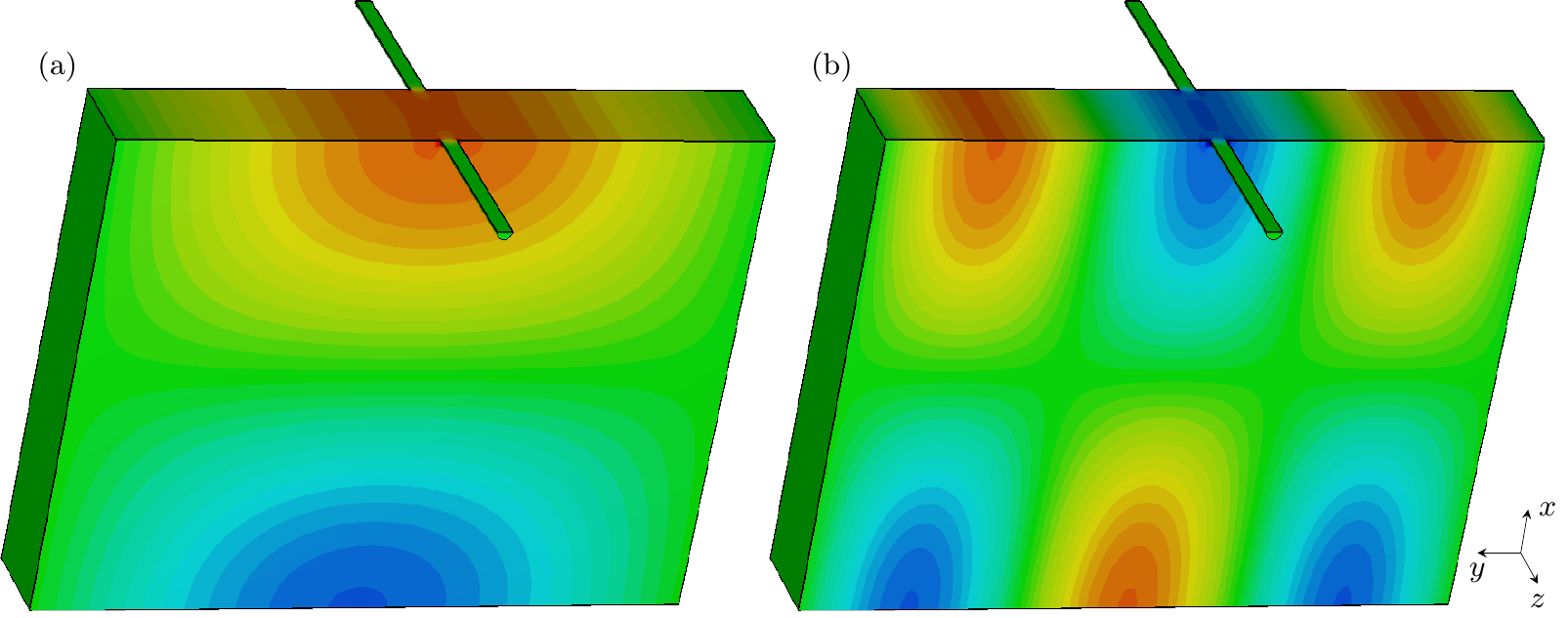}
	\caption{\label{fig:choppermodes}Cross section of the (a) fundamental and (b) higher harmonic chopping mode cavity. Colors represent the relative field strength of the magnetic field in the $y$-direction with red positive and blue negative, normalized to the maximum field strength. Cavity dimensions in mm are $379.1\times244.8\times36.1$.}
\end{figure*}
\begin{figure*}[t]
	\includegraphics[width=17.2cm]{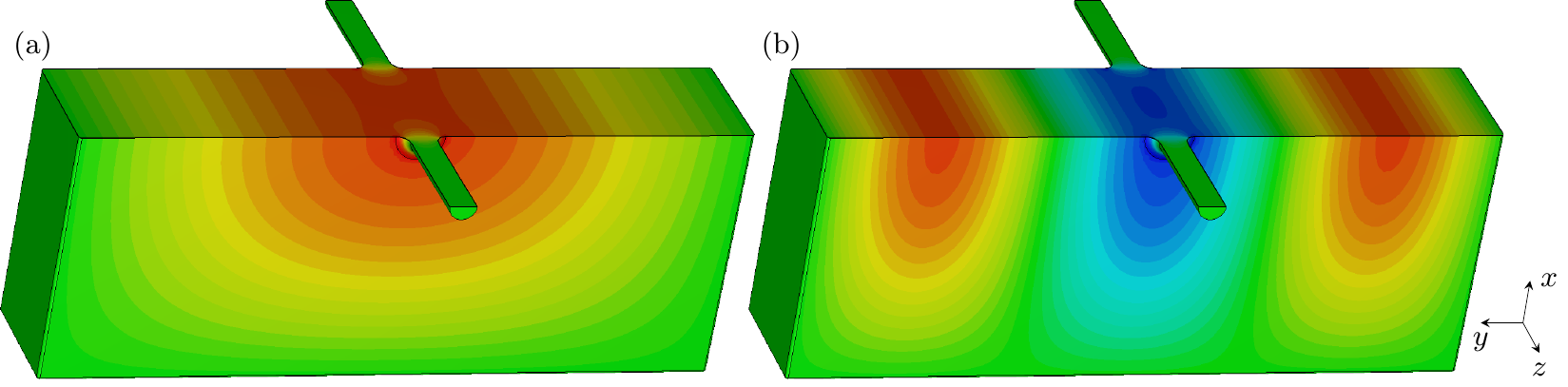}
	\caption{\label{fig:bunchermodes}Cross section of the (a) fundamental and (b) higher harmonic compression mode cavity. Colors represent the relative field strength of the electric field in the $z$-direction with red positive and blue negative, normalized to the maximum field strength. Cavity dimensions in mm are $189.6\times244.78\times47.3$.}
\end{figure*}
In order to use the RF cavities in \textsc{gpt}, a description of the 3D electromagnetic fields is required. 
To generate the required fields, a preliminary EM design was made for the rectangular chopping and compression cavities, shown in Figs.~\ref{fig:choppermodes} and \ref{fig:bunchermodes} respectively. The cavity dimensions agree to within $1$~mm with those calculated with Eqs.~(\ref{eq:harmratio}) and (\ref{eq:harmratio2}) and have a length of $36.1$~mm and $47.1$~mm as was found to be optimal in the simulations.\\ 
This design has not yet been optimized for power consumption. Yet as Sec.~\ref{subsec:cavfeas} shows, the power requirement to operate these at the desired field strengths is already within the capabilities of solid-state amplifiers.
\bibliography{bib_ZE10148_2}

\providecommand{\noopsort}[1]{}\providecommand{\singleletter}[1]{#1}%
\begin{thebibliography}{26}%
\makeatletter
\providecommand \@ifxundefined [1]{%
 \@ifx{#1\undefined}
}%
\providecommand \@ifnum [1]{%
 \ifnum #1\expandafter \@firstoftwo
 \else \expandafter \@secondoftwo
 \fi
}%
\providecommand \@ifx [1]{%
 \ifx #1\expandafter \@firstoftwo
 \else \expandafter \@secondoftwo
 \fi
}%
\providecommand \natexlab [1]{#1}%
\providecommand \enquote  [1]{``#1''}%
\providecommand \bibnamefont  [1]{#1}%
\providecommand \bibfnamefont [1]{#1}%
\providecommand \citenamefont [1]{#1}%
\providecommand \href@noop [0]{\@secondoftwo}%
\providecommand \href [0]{\begingroup \@sanitize@url \@href}%
\providecommand \@href[1]{\@@startlink{#1}\@@href}%
\providecommand \@@href[1]{\endgroup#1\@@endlink}%
\providecommand \@sanitize@url [0]{\catcode `\\12\catcode `\$12\catcode
  `\&12\catcode `\#12\catcode `\^12\catcode `\_12\catcode `\%12\relax}%
\providecommand \@@startlink[1]{}%
\providecommand \@@endlink[0]{}%
\providecommand \url  [0]{\begingroup\@sanitize@url \@url }%
\providecommand \@url [1]{\endgroup\@href {#1}{\urlprefix }}%
\providecommand \urlprefix  [0]{URL }%
\providecommand \Eprint [0]{\href }%
\providecommand \doibase [0]{https://doi.org/}%
\providecommand \selectlanguage [0]{\@gobble}%
\providecommand \bibinfo  [0]{\@secondoftwo}%
\providecommand \bibfield  [0]{\@secondoftwo}%
\providecommand \translation [1]{[#1]}%
\providecommand \BibitemOpen [0]{}%
\providecommand \bibitemStop [0]{}%
\providecommand \bibitemNoStop [0]{.\EOS\space}%
\providecommand \EOS [0]{\spacefactor3000\relax}%
\providecommand \BibitemShut  [1]{\csname bibitem#1\endcsname}%
\let\auto@bib@innerbib\@empty
\bibitem [{\citenamefont {Graves}\ \emph {et~al.}(2014)\citenamefont {Graves},
  \citenamefont {Bessuille}, \citenamefont {Brown}, \citenamefont {Carbajo},
  \citenamefont {Dolgashev}, \citenamefont {Hong}, \citenamefont {Ihloff},
  \citenamefont {Khaykovich}, \citenamefont {Lin}, \citenamefont {Murari},
  \citenamefont {Nanni}, \citenamefont {Resta}, \citenamefont {Tantawi},
  \citenamefont {Zapata}, \citenamefont {K\"artner},\ and\ \citenamefont
  {Moncton}}]{Graves2014}%
  \BibitemOpen
  \bibfield  {author} {\bibinfo {author} {\bibfnamefont {W.~S.}\ \bibnamefont
  {Graves}}, \bibinfo {author} {\bibfnamefont {J.}~\bibnamefont {Bessuille}},
  \bibinfo {author} {\bibfnamefont {P.}~\bibnamefont {Brown}}, \bibinfo
  {author} {\bibfnamefont {S.}~\bibnamefont {Carbajo}}, \bibinfo {author}
  {\bibfnamefont {V.}~\bibnamefont {Dolgashev}}, \bibinfo {author}
  {\bibfnamefont {K.-H.}\ \bibnamefont {Hong}}, \bibinfo {author}
  {\bibfnamefont {E.}~\bibnamefont {Ihloff}}, \bibinfo {author} {\bibfnamefont
  {B.}~\bibnamefont {Khaykovich}}, \bibinfo {author} {\bibfnamefont
  {H.}~\bibnamefont {Lin}}, \bibinfo {author} {\bibfnamefont {K.}~\bibnamefont
  {Murari}}, \bibinfo {author} {\bibfnamefont {E.~A.}\ \bibnamefont {Nanni}},
  \bibinfo {author} {\bibfnamefont {G.}~\bibnamefont {Resta}}, \bibinfo
  {author} {\bibfnamefont {S.}~\bibnamefont {Tantawi}}, \bibinfo {author}
  {\bibfnamefont {L.~E.}\ \bibnamefont {Zapata}}, \bibinfo {author}
  {\bibfnamefont {F.~X.}\ \bibnamefont {K\"artner}},\ and\ \bibinfo {author}
  {\bibfnamefont {D.~E.}\ \bibnamefont {Moncton}},\ }\bibfield  {title}
  {\bibinfo {title} {Compact x-ray source based on burst-mode inverse compton
  scattering at 100 khz},\ }\href
  {https://doi.org/10.1103/PhysRevSTAB.17.120701} {\bibfield  {journal}
  {\bibinfo  {journal} {Phys. Rev. ST Accel. Beams}\ }\textbf {\bibinfo
  {volume} {17}},\ \bibinfo {pages} {120701} (\bibinfo {year}
  {2014})}\BibitemShut {NoStop}%
\bibitem [{\citenamefont {Batchelor}\ \emph {et~al.}(1988)\citenamefont
  {Batchelor}, \citenamefont {Kirk}, \citenamefont {Sheehan}, \citenamefont
  {Woodle},\ and\ \citenamefont {McDonald}}]{Batchelor1988}%
  \BibitemOpen
  \bibfield  {author} {\bibinfo {author} {\bibfnamefont {K.}~\bibnamefont
  {Batchelor}}, \bibinfo {author} {\bibfnamefont {H.}~\bibnamefont {Kirk}},
  \bibinfo {author} {\bibfnamefont {J.}~\bibnamefont {Sheehan}}, \bibinfo
  {author} {\bibfnamefont {M.}~\bibnamefont {Woodle}},\ and\ \bibinfo {author}
  {\bibfnamefont {K.}~\bibnamefont {McDonald}},\ }\bibfield  {title} {\bibinfo
  {title} {{Development of a High Brightness Electron Gun for the Accelerator
  Test Facility at Brookhaven National Laboratory}},\ }\bibfield  {booktitle}
  {\emph {\bibinfo {booktitle} {{Particle accelerator. Proceedings, 1st EPAC
  Conference, Rome, Italy, June 7-11, 1988. Vol. 1, 2}}},\ }\href@noop {}
  {\bibfield  {journal} {\bibinfo  {journal} {Conf. Proc.}\ }\textbf {\bibinfo
  {volume} {C880607}},\ \bibinfo {pages} {954} (\bibinfo {year}
  {1988})}\BibitemShut {NoStop}%
\bibitem [{\citenamefont {Filippetto}\ \emph {et~al.}(2015)\citenamefont
  {Filippetto}, \citenamefont {Qian},\ and\ \citenamefont
  {Sannibale}}]{Filippetto2015}%
  \BibitemOpen
  \bibfield  {author} {\bibinfo {author} {\bibfnamefont {D.}~\bibnamefont
  {Filippetto}}, \bibinfo {author} {\bibfnamefont {H.}~\bibnamefont {Qian}},\
  and\ \bibinfo {author} {\bibfnamefont {F.}~\bibnamefont {Sannibale}},\
  }\bibfield  {title} {\bibinfo {title} {Cesium telluride cathodes for the next
  generation of high-average current high-brightness photoinjectors},\ }\href
  {https://doi.org/10.1063/1.4927700} {\bibfield  {journal} {\bibinfo
  {journal} {Applied Physics Letters}\ }\textbf {\bibinfo {volume} {107}},\
  \bibinfo {pages} {042104} (\bibinfo {year} {2015})},\ \Eprint
  {https://arxiv.org/abs/https://doi.org/10.1063/1.4927700}
  {https://doi.org/10.1063/1.4927700} \BibitemShut {NoStop}%
\bibitem [{\citenamefont {Dowell}\ \emph {et~al.}(2010)\citenamefont {Dowell},
  \citenamefont {Bazarov}, \citenamefont {Dunham}, \citenamefont {Harkay},
  \citenamefont {Hernandez-Garcia}, \citenamefont {Legg}, \citenamefont
  {Padmore}, \citenamefont {Rao}, \citenamefont {Smedley},\ and\ \citenamefont
  {Wan}}]{Dowel2010}%
  \BibitemOpen
  \bibfield  {author} {\bibinfo {author} {\bibfnamefont {D.}~\bibnamefont
  {Dowell}}, \bibinfo {author} {\bibfnamefont {I.}~\bibnamefont {Bazarov}},
  \bibinfo {author} {\bibfnamefont {B.}~\bibnamefont {Dunham}}, \bibinfo
  {author} {\bibfnamefont {K.}~\bibnamefont {Harkay}}, \bibinfo {author}
  {\bibfnamefont {C.}~\bibnamefont {Hernandez-Garcia}}, \bibinfo {author}
  {\bibfnamefont {R.}~\bibnamefont {Legg}}, \bibinfo {author} {\bibfnamefont
  {H.}~\bibnamefont {Padmore}}, \bibinfo {author} {\bibfnamefont
  {T.}~\bibnamefont {Rao}}, \bibinfo {author} {\bibfnamefont {J.}~\bibnamefont
  {Smedley}},\ and\ \bibinfo {author} {\bibfnamefont {W.}~\bibnamefont {Wan}},\
  }\bibfield  {title} {\bibinfo {title} {{Cathode R{\&}D for future light
  sources}},\ }\href
  {https://doi.org/https://doi.org/10.1016/j.nima.2010.03.104} {\bibfield
  {journal} {\bibinfo  {journal} {Nuclear Instruments and Methods in Physics
  Research Section A: Accelerators, Spectrometers, Detectors and Associated
  Equipment}\ }\textbf {\bibinfo {volume} {622}},\ \bibinfo {pages} {685 }
  (\bibinfo {year} {2010})}\BibitemShut {NoStop}%
\bibitem [{\citenamefont {Musumeci}\ \emph {et~al.}(2018)\citenamefont
  {Musumeci}, \citenamefont {Navarro}, \citenamefont {Rosenzweig},
  \citenamefont {Cultrera}, \citenamefont {Bazarov}, \citenamefont {Maxson},
  \citenamefont {Karkare},\ and\ \citenamefont {Padmore}}]{Musumeci2018}%
  \BibitemOpen
  \bibfield  {author} {\bibinfo {author} {\bibfnamefont {P.}~\bibnamefont
  {Musumeci}}, \bibinfo {author} {\bibfnamefont {J.~G.}\ \bibnamefont
  {Navarro}}, \bibinfo {author} {\bibfnamefont {J.}~\bibnamefont {Rosenzweig}},
  \bibinfo {author} {\bibfnamefont {L.}~\bibnamefont {Cultrera}}, \bibinfo
  {author} {\bibfnamefont {I.}~\bibnamefont {Bazarov}}, \bibinfo {author}
  {\bibfnamefont {J.}~\bibnamefont {Maxson}}, \bibinfo {author} {\bibfnamefont
  {S.}~\bibnamefont {Karkare}},\ and\ \bibinfo {author} {\bibfnamefont
  {H.}~\bibnamefont {Padmore}},\ }\bibfield  {title} {\bibinfo {title}
  {Advances in bright electron sources},\ }\href
  {https://doi.org/https://doi.org/10.1016/j.nima.2018.03.019} {\bibfield
  {journal} {\bibinfo  {journal} {Nuclear Instruments and Methods in Physics
  Research Section A: Accelerators, Spectrometers, Detectors and Associated
  Equipment}\ }\textbf {\bibinfo {volume} {907}},\ \bibinfo {pages} {209 }
  (\bibinfo {year} {2018})},\ \bibinfo {note} {advances in Instrumentation and
  Experimental Methods (Special Issue in Honour of Kai Siegbahn)}\BibitemShut
  {NoStop}%
\bibitem [{\citenamefont {Sannibale}\ \emph {et~al.}(2012)\citenamefont
  {Sannibale}, \citenamefont {Filippetto}, \citenamefont {Papadopoulos},
  \citenamefont {Staples}, \citenamefont {Wells}, \citenamefont {Bailey},
  \citenamefont {Baptiste}, \citenamefont {Corlett}, \citenamefont {Cork},
  \citenamefont {De~Santis}, \citenamefont {Dimaggio}, \citenamefont
  {Doolittle}, \citenamefont {Doyle}, \citenamefont {Feng}, \citenamefont
  {Garcia~Quintas}, \citenamefont {Huang}, \citenamefont {Huang}, \citenamefont
  {Kramasz}, \citenamefont {Kwiatkowski}, \citenamefont {Lellinger},
  \citenamefont {Moroz}, \citenamefont {Norum}, \citenamefont {Padmore},
  \citenamefont {Pappas}, \citenamefont {Portmann}, \citenamefont {Vecchione},
  \citenamefont {Vinco}, \citenamefont {Zolotorev},\ and\ \citenamefont
  {Zucca}}]{Sannibale2012}%
  \BibitemOpen
  \bibfield  {author} {\bibinfo {author} {\bibfnamefont {F.}~\bibnamefont
  {Sannibale}}, \bibinfo {author} {\bibfnamefont {D.}~\bibnamefont
  {Filippetto}}, \bibinfo {author} {\bibfnamefont {C.~F.}\ \bibnamefont
  {Papadopoulos}}, \bibinfo {author} {\bibfnamefont {J.}~\bibnamefont
  {Staples}}, \bibinfo {author} {\bibfnamefont {R.}~\bibnamefont {Wells}},
  \bibinfo {author} {\bibfnamefont {B.}~\bibnamefont {Bailey}}, \bibinfo
  {author} {\bibfnamefont {K.}~\bibnamefont {Baptiste}}, \bibinfo {author}
  {\bibfnamefont {J.}~\bibnamefont {Corlett}}, \bibinfo {author} {\bibfnamefont
  {C.}~\bibnamefont {Cork}}, \bibinfo {author} {\bibfnamefont {S.}~\bibnamefont
  {De~Santis}}, \bibinfo {author} {\bibfnamefont {S.}~\bibnamefont {Dimaggio}},
  \bibinfo {author} {\bibfnamefont {L.}~\bibnamefont {Doolittle}}, \bibinfo
  {author} {\bibfnamefont {J.}~\bibnamefont {Doyle}}, \bibinfo {author}
  {\bibfnamefont {J.}~\bibnamefont {Feng}}, \bibinfo {author} {\bibfnamefont
  {D.}~\bibnamefont {Garcia~Quintas}}, \bibinfo {author} {\bibfnamefont
  {G.}~\bibnamefont {Huang}}, \bibinfo {author} {\bibfnamefont
  {H.}~\bibnamefont {Huang}}, \bibinfo {author} {\bibfnamefont
  {T.}~\bibnamefont {Kramasz}}, \bibinfo {author} {\bibfnamefont
  {S.}~\bibnamefont {Kwiatkowski}}, \bibinfo {author} {\bibfnamefont
  {R.}~\bibnamefont {Lellinger}}, \bibinfo {author} {\bibfnamefont
  {V.}~\bibnamefont {Moroz}}, \bibinfo {author} {\bibfnamefont {W.~E.}\
  \bibnamefont {Norum}}, \bibinfo {author} {\bibfnamefont {H.}~\bibnamefont
  {Padmore}}, \bibinfo {author} {\bibfnamefont {C.}~\bibnamefont {Pappas}},
  \bibinfo {author} {\bibfnamefont {G.}~\bibnamefont {Portmann}}, \bibinfo
  {author} {\bibfnamefont {T.}~\bibnamefont {Vecchione}}, \bibinfo {author}
  {\bibfnamefont {M.}~\bibnamefont {Vinco}}, \bibinfo {author} {\bibfnamefont
  {M.}~\bibnamefont {Zolotorev}},\ and\ \bibinfo {author} {\bibfnamefont
  {F.}~\bibnamefont {Zucca}},\ }\bibfield  {title} {\bibinfo {title} {Advanced
  photoinjector experiment photogun commissioning results},\ }\href
  {https://doi.org/10.1103/PhysRevSTAB.15.103501} {\bibfield  {journal}
  {\bibinfo  {journal} {Phys. Rev. ST Accel. Beams}\ }\textbf {\bibinfo
  {volume} {15}},\ \bibinfo {pages} {103501} (\bibinfo {year}
  {2012})}\BibitemShut {NoStop}%
\bibitem [{\citenamefont {Vogel}\ \emph {et~al.}(2018)\citenamefont {Vogel}
  \emph {et~al.}}]{Vogel2018}%
  \BibitemOpen
  \bibfield  {author} {\bibinfo {author} {\bibfnamefont {E.}~\bibnamefont
  {Vogel}} \emph {et~al.},\ }\bibfield  {title} {\bibinfo {title} {{SRF Gun
  Development at DESY}},\ }in\ \href
  {https://doi.org/10.18429/JACoW-LINAC2018-MOPO037} {\emph {\bibinfo
  {booktitle} {{Proceedings, 29th International Linear Accelerator Conference
  (LINAC18): Beijing, China, September 16-21, 2018}}}}\ (\bibinfo {year}
  {2018})\ p.\ \bibinfo {pages} {MOPO037}\BibitemShut {NoStop}%
\bibitem [{\citenamefont {Bartnik}\ \emph {et~al.}(2015)\citenamefont
  {Bartnik}, \citenamefont {Gulliford}, \citenamefont {Bazarov}, \citenamefont
  {Cultera},\ and\ \citenamefont {Dunham}}]{Bartnik2015}%
  \BibitemOpen
  \bibfield  {author} {\bibinfo {author} {\bibfnamefont {A.}~\bibnamefont
  {Bartnik}}, \bibinfo {author} {\bibfnamefont {C.}~\bibnamefont {Gulliford}},
  \bibinfo {author} {\bibfnamefont {I.}~\bibnamefont {Bazarov}}, \bibinfo
  {author} {\bibfnamefont {L.}~\bibnamefont {Cultera}},\ and\ \bibinfo {author}
  {\bibfnamefont {B.}~\bibnamefont {Dunham}},\ }\bibfield  {title} {\bibinfo
  {title} {Operational experience with nanocoulomb bunch charges in the
  {C}ornell photoinjector},\ }\href
  {https://doi.org/10.1103/PhysRevSTAB.18.083401} {\bibfield  {journal}
  {\bibinfo  {journal} {Phys. Rev. ST Accel. Beams}\ }\textbf {\bibinfo
  {volume} {18}},\ \bibinfo {pages} {083401} (\bibinfo {year}
  {2015})}\BibitemShut {NoStop}%
\bibitem [{\citenamefont {Gulliford}\ \emph {et~al.}(2013)\citenamefont
  {Gulliford}, \citenamefont {Bartnik}, \citenamefont {Bazarov}, \citenamefont
  {Cultrera}, \citenamefont {Dobbins}, \citenamefont {Dunham}, \citenamefont
  {Gonzalez}, \citenamefont {Karkare}, \citenamefont {Lee}, \citenamefont {Li}
  \emph {et~al.}}]{Gulliford2013}%
  \BibitemOpen
  \bibfield  {author} {\bibinfo {author} {\bibfnamefont {C.}~\bibnamefont
  {Gulliford}}, \bibinfo {author} {\bibfnamefont {A.}~\bibnamefont {Bartnik}},
  \bibinfo {author} {\bibfnamefont {I.}~\bibnamefont {Bazarov}}, \bibinfo
  {author} {\bibfnamefont {L.}~\bibnamefont {Cultrera}}, \bibinfo {author}
  {\bibfnamefont {J.}~\bibnamefont {Dobbins}}, \bibinfo {author} {\bibfnamefont
  {B.}~\bibnamefont {Dunham}}, \bibinfo {author} {\bibfnamefont
  {F.}~\bibnamefont {Gonzalez}}, \bibinfo {author} {\bibfnamefont
  {S.}~\bibnamefont {Karkare}}, \bibinfo {author} {\bibfnamefont
  {H.}~\bibnamefont {Lee}}, \bibinfo {author} {\bibfnamefont {H.}~\bibnamefont
  {Li}}, \emph {et~al.},\ }\bibfield  {title} {\bibinfo {title} {Demonstration
  of low emittance in the {C}ornell energy recovery linac injector prototype},\
  }\href {https://doi.org/10.1103/PhysRevSTAB.16.073401} {\bibfield  {journal}
  {\bibinfo  {journal} {Phys. Rev. ST Accel. Beams}\ }\textbf {\bibinfo
  {volume} {16}},\ \bibinfo {pages} {073401} (\bibinfo {year}
  {2013})}\BibitemShut {NoStop}%
\bibitem [{\citenamefont {Jenkins}(1969)}]{Jenkins1969}%
  \BibitemOpen
  \bibfield  {author} {\bibinfo {author} {\bibfnamefont {R.}~\bibnamefont
  {Jenkins}},\ }\bibfield  {title} {\bibinfo {title} {A review of thermionic
  cathodes},\ }\href
  {https://doi.org/https://doi.org/10.1016/S0042-207X(69)80077-1} {\bibfield
  {journal} {\bibinfo  {journal} {Vacuum}\ }\textbf {\bibinfo {volume} {19}},\
  \bibinfo {pages} {353 } (\bibinfo {year} {1969})}\BibitemShut {NoStop}%
\bibitem [{\citenamefont {Togawa}\ \emph {et~al.}(2007)\citenamefont {Togawa},
  \citenamefont {Shintake}, \citenamefont {Inagaki}, \citenamefont {Onoe},
  \citenamefont {Tanaka}, \citenamefont {Baba},\ and\ \citenamefont
  {Matsumoto}}]{Togawa2007}%
  \BibitemOpen
  \bibfield  {author} {\bibinfo {author} {\bibfnamefont {K.}~\bibnamefont
  {Togawa}}, \bibinfo {author} {\bibfnamefont {T.}~\bibnamefont {Shintake}},
  \bibinfo {author} {\bibfnamefont {T.}~\bibnamefont {Inagaki}}, \bibinfo
  {author} {\bibfnamefont {K.}~\bibnamefont {Onoe}}, \bibinfo {author}
  {\bibfnamefont {T.}~\bibnamefont {Tanaka}}, \bibinfo {author} {\bibfnamefont
  {H.}~\bibnamefont {Baba}},\ and\ \bibinfo {author} {\bibfnamefont
  {H.}~\bibnamefont {Matsumoto}},\ }\bibfield  {title} {\bibinfo {title}
  {{${\mathrm{CeB}}_{6}$ electron gun for low-emittance injector}},\ }\href
  {https://doi.org/10.1103/PhysRevSTAB.10.020703} {\bibfield  {journal}
  {\bibinfo  {journal} {Phys. Rev. ST Accel. Beams}\ }\textbf {\bibinfo
  {volume} {10}},\ \bibinfo {pages} {020703} (\bibinfo {year}
  {2007})}\BibitemShut {NoStop}%
\bibitem [{\citenamefont {Abbott}\ \emph {et~al.}(1994)\citenamefont {Abbott},
  \citenamefont {Benson}, \citenamefont {Crofford}, \citenamefont {Douglas},
  \citenamefont {Gonzales},\ and\ \citenamefont {Kazimi}}]{Abbott1994}%
  \BibitemOpen
  \bibfield  {author} {\bibinfo {author} {\bibfnamefont {R.}~\bibnamefont
  {Abbott}}, \bibinfo {author} {\bibfnamefont {S.}~\bibnamefont {Benson}},
  \bibinfo {author} {\bibfnamefont {M.}~\bibnamefont {Crofford}}, \bibinfo
  {author} {\bibfnamefont {D.}~\bibnamefont {Douglas}}, \bibinfo {author}
  {\bibfnamefont {R.}~\bibnamefont {Gonzales}},\ and\ \bibinfo {author}
  {\bibfnamefont {R.}~\bibnamefont {Kazimi}},\ }\bibfield  {title} {\bibinfo
  {title} {Design, commissioning, and operation of the upgraded {CEBAF}
  injector},\ }in\ \href@noop {} {\emph {\bibinfo {booktitle} {1994
  International Linac Conference}}},\ \bibinfo {editor} {edited by\ \bibinfo
  {editor} {\bibfnamefont {K.}~\bibnamefont {Takata}}, \bibinfo {editor}
  {\bibfnamefont {K.}~\bibnamefont {Nakahara}},\ and\ \bibinfo {editor}
  {\bibfnamefont {Y.}~\bibnamefont {Yamazaki}}}\ (\bibinfo {address} {Tsukuba,
  Japan},\ \bibinfo {year} {1994})\ pp.\ \bibinfo {pages}
  {777--779}\BibitemShut {NoStop}%
\bibitem [{\citenamefont {{Vretenar}}(2013)}]{Vretenar2013}%
  \BibitemOpen
  \bibfield  {author} {\bibinfo {author} {\bibfnamefont {M.}~\bibnamefont
  {{Vretenar}}},\ }\bibfield  {title} {\bibinfo {title} {{Linear
  accelerators}},\ }\href@noop {} {\bibfield  {journal} {\bibinfo  {journal}
  {arXiv e-prints}\ } (\bibinfo {year} {2013})},\ \Eprint
  {https://arxiv.org/abs/1303.6766} {arXiv:1303.6766 [physics.acc-ph]}
  \BibitemShut {NoStop}%
\bibitem [{\citenamefont {Lassise}\ \emph {et~al.}(2012)\citenamefont
  {Lassise}, \citenamefont {Mutsaers},\ and\ \citenamefont
  {Luiten}}]{Lassise2012}%
  \BibitemOpen
  \bibfield  {author} {\bibinfo {author} {\bibfnamefont {A.}~\bibnamefont
  {Lassise}}, \bibinfo {author} {\bibfnamefont {P.}~\bibnamefont {Mutsaers}},\
  and\ \bibinfo {author} {\bibfnamefont {O.}~\bibnamefont {Luiten}},\
  }\bibfield  {title} {\bibinfo {title} {{Compact, low power radio frequency
  cavity for femtosecond electron microscopy}},\ }\href
  {https://doi.org/10.1063/1.3703314} {\bibfield  {journal} {\bibinfo
  {journal} {Review of Scientific Instruments}\ }\textbf {\bibinfo {volume}
  {83}},\ \bibinfo {pages} {1} (\bibinfo {year} {2012})}\BibitemShut {NoStop}%
\bibitem [{\citenamefont {Dushman}(1923)}]{Dushman1923}%
  \BibitemOpen
  \bibfield  {author} {\bibinfo {author} {\bibfnamefont {S.}~\bibnamefont
  {Dushman}},\ }\bibfield  {title} {\bibinfo {title} {{Electron Emission from
  Metals as a Function of Temperature}},\ }\href
  {https://doi.org/10.1103/PhysRev.21.623} {\bibfield  {journal} {\bibinfo
  {journal} {Phys. Rev.}\ }\textbf {\bibinfo {volume} {21}},\ \bibinfo {pages}
  {623} (\bibinfo {year} {1923})}\BibitemShut {NoStop}%
\bibitem [{\citenamefont {Orloff}(2009)}]{Orloff2009}%
  \BibitemOpen
  \bibfield  {author} {\bibinfo {author} {\bibfnamefont {J.}~\bibnamefont
  {Orloff}},\ }\href@noop {} {\emph {\bibinfo {title} {Handbook of charged
  particle optics}}}\ (\bibinfo  {publisher} {CRC press},\ \bibinfo {year}
  {2009})\BibitemShut {NoStop}%
\bibitem [{\citenamefont {Lafferty}(1951)}]{Lafferty1951}%
  \BibitemOpen
  \bibfield  {author} {\bibinfo {author} {\bibfnamefont {J.~M.}\ \bibnamefont
  {Lafferty}},\ }\bibfield  {title} {\bibinfo {title} {Boride cathodes},\
  }\href {https://doi.org/10.1063/1.1699946} {\bibfield  {journal} {\bibinfo
  {journal} {Journal of Applied Physics}\ }\textbf {\bibinfo {volume} {22}},\
  \bibinfo {pages} {299} (\bibinfo {year} {1951})},\ \Eprint
  {https://arxiv.org/abs/https://doi.org/10.1063/1.1699946}
  {https://doi.org/10.1063/1.1699946} \BibitemShut {NoStop}%
\bibitem [{\citenamefont {Pelletier}\ and\ \citenamefont
  {Pomot}(1979)}]{Pelletier1979}%
  \BibitemOpen
  \bibfield  {author} {\bibinfo {author} {\bibfnamefont {J.}~\bibnamefont
  {Pelletier}}\ and\ \bibinfo {author} {\bibfnamefont {C.}~\bibnamefont
  {Pomot}},\ }\bibfield  {title} {\bibinfo {title} {Work function of sintered
  lanthanum hexaboride},\ }\href {https://doi.org/10.1063/1.90769} {\bibfield
  {journal} {\bibinfo  {journal} {Applied Physics Letters}\ }\textbf {\bibinfo
  {volume} {34}},\ \bibinfo {pages} {249} (\bibinfo {year} {1979})},\ \Eprint
  {https://arxiv.org/abs/https://doi.org/10.1063/1.90769}
  {https://doi.org/10.1063/1.90769} \BibitemShut {NoStop}%
\bibitem [{\citenamefont {Floettmann}(2003)}]{Floettmann2003}%
  \BibitemOpen
  \bibfield  {author} {\bibinfo {author} {\bibfnamefont {K.}~\bibnamefont
  {Floettmann}},\ }\bibfield  {title} {\bibinfo {title} {Some basic features of
  the beam emittance},\ }\href {https://doi.org/10.1103/PhysRevSTAB.6.034202}
  {\bibfield  {journal} {\bibinfo  {journal} {Phys. Rev. ST Accel. Beams}\
  }\textbf {\bibinfo {volume} {6}},\ \bibinfo {pages} {034202} (\bibinfo {year}
  {2003})}\BibitemShut {NoStop}%
\bibitem [{\citenamefont {Miltchev}(2005)}]{Miltchev2005}%
  \BibitemOpen
  \bibfield  {author} {\bibinfo {author} {\bibfnamefont {V.}~\bibnamefont
  {Miltchev}},\ }\bibfield  {title} {\bibinfo {title} {Modelling the transverse
  phase space and core emittance studies at pitz},\ }in\ \href@noop {} {\emph
  {\bibinfo {booktitle} {27th International Conference on Free Electron
  Lasers}}},\ \bibinfo {editor} {edited by\ \bibinfo {editor} {\bibfnamefont
  {J.}~\bibnamefont {Galayda}}\ and\ \bibinfo {editor} {\bibfnamefont
  {I.}~\bibnamefont {Lindau}}}\ (\bibinfo {address} {Palo Alto, CA, USA},\
  \bibinfo {year} {2005})\ pp.\ \bibinfo {pages} {556--559}\BibitemShut
  {NoStop}%
\bibitem [{\citenamefont {Child}(1911)}]{Child1911}%
  \BibitemOpen
  \bibfield  {author} {\bibinfo {author} {\bibfnamefont {C.~D.}\ \bibnamefont
  {Child}},\ }\bibfield  {title} {\bibinfo {title} {{Discharge From Hot CaO}},\
  }\href {https://doi.org/10.1103/PhysRevSeriesI.32.492} {\bibfield  {journal}
  {\bibinfo  {journal} {Phys. Rev. (Series I)}\ }\textbf {\bibinfo {volume}
  {32}},\ \bibinfo {pages} {492} (\bibinfo {year} {1911})}\BibitemShut
  {NoStop}%
\bibitem [{CST()}]{CST}%
  \BibitemOpen
  \href@noop {} {}\bibinfo {howpublished}
  {\url{https://www.cst.com}}\BibitemShut {NoStop}%
\bibitem [{\citenamefont {Palmer}\ \emph {et~al.}(1997)\citenamefont {Palmer},
  \citenamefont {Wang}, \citenamefont {Ben-Zvi}, \citenamefont {Miller},\ and\
  \citenamefont {Skaritka}}]{Palmer1997}%
  \BibitemOpen
  \bibfield  {author} {\bibinfo {author} {\bibfnamefont {D.~T.}\ \bibnamefont
  {Palmer}}, \bibinfo {author} {\bibfnamefont {X.~J.}\ \bibnamefont {Wang}},
  \bibinfo {author} {\bibfnamefont {I.}~\bibnamefont {Ben-Zvi}}, \bibinfo
  {author} {\bibfnamefont {R.~H.}\ \bibnamefont {Miller}},\ and\ \bibinfo
  {author} {\bibfnamefont {J.}~\bibnamefont {Skaritka}},\ }\bibfield  {title}
  {\bibinfo {title} {{Experimental results of a single emittance compensation
  solenoidal magnet}},\ }\bibfield  {booktitle} {\emph {\bibinfo {booktitle}
  {{17th IEEE Particle Accelerator Conference (PAC 97): Accelerator Science,
  Technology and Applications Vancouver, British Columbia, Canada, May 12-16,
  1997}}},\ }\href@noop {} {\bibfield  {journal} {\bibinfo  {journal} {Conf.
  Proc.}\ }\textbf {\bibinfo {volume} {C970512}},\ \bibinfo {pages} {2843}
  (\bibinfo {year} {1997})}\BibitemShut {NoStop}%
\bibitem [{\citenamefont {Lassise}(2012)}]{Lassise2012thesis}%
  \BibitemOpen
  \bibfield  {author} {\bibinfo {author} {\bibfnamefont {A.}~\bibnamefont
  {Lassise}},\ }\emph {\bibinfo {title} {{Miniaturized RF technology for
  femtosecond electron microscopy}}},\ \href {https://doi.org/10.6100/IR739203}
  {Ph.D. thesis},\ \bibinfo  {school} {Eindhoven University of Technology}
  (\bibinfo {year} {2012})\BibitemShut {NoStop}%
\bibitem [{GPT()}]{GPT}%
  \BibitemOpen
  \href@noop {} {}\bibinfo {howpublished}
  {\url{http://www.pulsar.nl/gpt}}\BibitemShut {NoStop}%
\bibitem [{\citenamefont {Verhoeven}\ \emph {et~al.}(2018)\citenamefont
  {Verhoeven}, \citenamefont {van Rens}, \citenamefont {Kemper}, \citenamefont
  {Rietman}, \citenamefont {van Doorn}, \citenamefont {Koole}, \citenamefont
  {Kieft}, \citenamefont {Mutsaers},\ and\ \citenamefont
  {Luiten}}]{Verhoeven2018}%
  \BibitemOpen
  \bibfield  {author} {\bibinfo {author} {\bibfnamefont {W.}~\bibnamefont
  {Verhoeven}}, \bibinfo {author} {\bibfnamefont {J.~F.~M.}\ \bibnamefont {van
  Rens}}, \bibinfo {author} {\bibfnamefont {A.~H.}\ \bibnamefont {Kemper}},
  \bibinfo {author} {\bibfnamefont {E.~H.}\ \bibnamefont {Rietman}}, \bibinfo
  {author} {\bibfnamefont {H.~A.}\ \bibnamefont {van Doorn}}, \bibinfo {author}
  {\bibfnamefont {I.}~\bibnamefont {Koole}}, \bibinfo {author} {\bibfnamefont
  {E.~R.}\ \bibnamefont {Kieft}}, \bibinfo {author} {\bibfnamefont {P.~H.~A.}\
  \bibnamefont {Mutsaers}},\ and\ \bibinfo {author} {\bibfnamefont {O.~J.}\
  \bibnamefont {Luiten}},\ }\href@noop {} {\bibinfo {title} {Design and
  characterization of dielectric filled {TM}$_{110}$ microwave cavities for
  ultrafast electron microscopy}} (\bibinfo {year} {2018}),\ \Eprint
  {https://arxiv.org/abs/arXiv:1811.02243} {arXiv:1811.02243} \BibitemShut
  {NoStop}%
\end{thebibliography}%
\end{document}